\documentclass{aa}
\usepackage[varg]{txfonts}
\usepackage{bm}
\usepackage{color}
\usepackage{natbib}
\usepackage{amssymb}
\usepackage{ulem}

\bibpunct{(}{)}{;}{a}{}{,} 

\newcommand{\rmd }{{\rm d}}  
\newcommand{\rmi }{{\rm i}}  

\newcommand{\taus}{{\tau_{\rm s}}}

\newcommand{\stokes}{{\rm St}}

\newcommand{\adi}{ARDI}

\begin{document}

\title{Resonant Drag Instabilities for Polydisperse Dust}
\subtitle{II. The streaming and settling instabilities}

\author{
Sijme-Jan Paardekooper\inst{1}
\and Hossam Aly\inst{1}
}

\institute{Faculty of Aerospace Engineering, Delft University of Technology, Delft,The Netherlands}

\date{Received *** / Accepted ***}

\abstract {Dust grains embedded in gas flow give rise to a class of hydrodynamic instabilities, called resonant drag instabilities, some of which are thought to be important during the process of planet formation. These instabilities have predominantly been studied for single grain sizes, in which case they are found to grow fast. Nonlinear simulations indicate that strong dust overdensities can form, aiding the formation of planetesimals. In reality, however, there is going to be a distribution of dust sizes, which potentially has important consequences.} 
{We study two different resonant drag instabilities, the streaming instability and the settling instability, taking into account a continuous spectrum of grain sizes, to determine whether these instabilities survive in the polydisperse regime and how the resulting growth rates compare to the monodisperse case.} 
{We solve the linear equations for a polydisperse fluid in an unstratified shearing box to recover the streaming instability and, for approximate stratification, the settling instability, in all cases focusing on small dust-to-gas ratios. } 
{Size distributions of realistic width turn the singular perturbation of the monodisperse limit into a regular perturbation due to the fact that the backreaction on the gas involves an integration over the resonance. The contribution of the resonance to the integral can be negative, as in the case of the streaming instability, which as a result does not survive in the polydisperse regime, or positive, which is the case in the settling instability. The latter therefore has a polydisperse counterpart, with growth rates that can be comparable to the monodisperse case.} 
{Wide size distributions in almost all cases remove the resonant nature of drag instabilities. This can lead to reduced growth, as is the case in large parts of parameter space for the settling instability, or complete stabilisation, as is the case for the streaming instability.}


\maketitle 

\section{Introduction}

Astrophysical fluids are often a mixture of gas and solid particles or \emph{dust}, at a canonical dust-to-gas ratio of $1$ $\%$ \citep[see e.g.][]{2007ApJ...663..866D}. Since different forces act on gas and dust (e.g. gas pressure only on gas, radiation pressure only on dust), the two components often have different equilibrium velocities. In many cases, such a situation where gas and dust drift relatively to each other is hydrodynamically unstable. A wide class of such instabilities are the so-called resonant drag instabilities \citep[RDIs, see][]{2018ApJ...856L..15S,2018MNRAS.477.5011S}.

In the analysis of RDIs, the dust-to-gas ratio $\mu$ is taken to be a small parameter. In the absence of dust, the gas can sustain a set of waves, depending on the relevant physics (e.g. sound waves, buoyancy waves, magnetosonic waves, etc). If the gas wave speed matches the drift speed of the dust, the dust induces a particularly strong reaction to the system, often resulting in exponentially growing perturbations \citep{2018MNRAS.477.5011S}. Matching the wave and drift speeds only happens at a specific wavelength (if the wave speed depends on wavelength) and dust size (if drift speed depends on size), hence the resonant nature of the instabilities. 

One can classify RDIs according to the associated gas wave. In a companion paper \citep[][hereafter Paper I]{paper1}, we studied the acoustic drag instability, which relies on a gas sound wave going unstable due to dust loading. If the gas is magnetized, or stably stratified, different waves are available to create an RDI \citep{2018MNRAS.477.5011S,2020MNRAS.496.2123H}. In this paper, we focus on two related instabilities that are relevant for the formation of planets, as they can occur in protoplanetary discs. They are the streaming instability \citep[SI,][]{2005ApJ...620..459Y} and the dust settling instability \citep[DSI,][]{2018MNRAS.477.5011S,2020MNRAS.497.2715K}. In both cases, the underlying gas wave is an inertial wave \citep{2019MNRAS.489.3850Z}.

The SI has found its way into planet formation theory because it can potentially solve the problem of the formation of planetesimals by creating overdense dust clumps that collapse under their own gravity \citep[for a recent review see][]{2023ASPC..534..465L}. This has always been a problematic part of planet formation theory, because rapid inward dust drift threatens to move all planetary building blocks into the central star \citep[e.g.][]{1977MNRAS.180...57W}. The SI is an elegant solution to this problem, as it actually uses the drift to trigger the RDI and build bodies large enough (i.e. km-sized planetesimals) in a short time scale, so that they are safe from drifting too far in.

Since dust drift usually depends on dust size, because smaller particles are more tightly coupled to the gas, extending the RDI formalism to incorporate a distribution of dust sizes is not straightforward. For a given gas wave, only a single particle size will drift at the resonant speed, but if the other sizes are passive one might hope that the feedback loop leading to the RDI \citep[see][]{2024MNRAS.tmp...61M} can survive. Indeed, some simulations 
indicate that the clumping induced by the SI is not very sensitive to having a size distribution \citep{2021A&A...653A..14S,2023MNRAS.526.1757R}. However, this turns out not to be true for the linear phase, at least in the case where vertical stratification is neglected \citep[as was done in the original work of][]{2005ApJ...620..459Y}.

The effect of a dust size distribution on the linear SI was investigated by \cite{2019ApJ...878L..30K}, who found that for a discrete number of different particle sizes, in many regions of parameter space the growth rate of the SI keeps decreasing with the number of particle sizes considered. It was shown in \cite{2020MNRAS.499.4223P} that in the limit of a continuous size distribution, the SI indeed ceases to exist except at extremely small wavelengths at dust-to-gas ratios much larger than unity \citep[see also][]{2021MNRAS.501..467Z}. These small scales are particularly prone to viscous damping, so that in addition a top-heavy size distribution is needed for the linear SI to survive \citep{2021MNRAS.502.1469M}. 

The nonlinear evolution of the SI with multiple dust sizes was studied in \cite{2021MNRAS.508.5538Y}, who found that one either needs dust to gas ratio $\mu>1$ or maximum stopping time $\taus > 1$ to get significant activity in the nonlinear regime. It should be noted again that these results, like the original SI analysis \citep{2005ApJ...620..459Y}, neglect vertical stratification in the disc, and that the SI may well have a different character in the stratified case \citep{2021ApJ...907...64L}, and react differently to a size distribution, at least in the nonlinear regime \citep[see e.g.][]{2021A&A...653A..14S,2023MNRAS.526.1757R}.

The settling instability is a more recently discovered RDI, also reliant on inertial waves, but taking into account vertical dust drift, i.e. settling \citep{2018MNRAS.477.5011S}. In \cite{2020MNRAS.497.2715K}, it was reported that this RDI is far less sensitive to a size distribution, and able to maintain strong growth rates even for wide size distributions. This raises a question as to which RDIs can survive in the presence of a dust size distribution, and therefore contribute to for example planetesimal formation \citep[e.g.][]{2007Natur.448.1022J}. In a companion paper, we showed that there exists a polydisperse version of the acoustic drag instability, but only for large wave lengths, which reduces the maximum growth rate significantly. In this paper, we focus on polydisperse versions of both the streaming and settling instability, in the regime where $\mu \ll 1$ to make contact with monodisperse RDI results. We will focus on whether these exist in a polydisperse context, and, if so, how the growth rates compare to the monodisperse case. 

The plan of this paper is as follows. We introduce the basic equations in section \ref{sec:eq_general}. We present the results on the streaming and settling instability in sections \ref{sec:stream} and \ref{sec:settling}, respectively, and we conclude in section \ref{sec:conclusion}.

\section{Governing equations}
\label{sec:eq_general}

The equations governing a mixture of gas and dust, where the dust component consists of a continuum of sizes, were presented in \citet[][see also Paper I]{2021MNRAS.502.1579P}:
\begin{align}
\partial_t\sigma 
+ \nabla\cdot(\sigma {\bf{u}} )=& 0,\label{eq:dustcont}\\
\partial_t{\bf{u}}
+ ({\bf{u}}\cdot\nabla){\bf{u}}
=&
\bm{\alpha}_{\rm d}  - \frac{{\bf{u}} - {\bf{v}_{\rm g}}}{\taus},\label{eq:dustmom}
\end{align}
with ${\bf u}$ the size-dependent dust velocity, and ${\bf{v}_{\rm g}}$ the gas velocity. The quantity 
$\sigma$ is the size density \citep{2020MNRAS.499.4223P}, defined such that
\begin{align}
    \rho_{\rm d}=\int \sigma \rmd \taus,
\end{align}
with $\rho_{\rm d}$ the dust density and $\taus$ the stopping time, which is a proxy for particle size. As in Paper I, we take the simplest drag law of the form $({\bf{u}}-{\bf{v}_{\rm g}})/\taus$ to couple gas and dust, with constant stopping time $\taus$. Any additional accelerations acting on the dust are contained in $\bm{\alpha}_{\rm d}$, which we assume to depend on ${\bf{u}}$ only.

The gas component, similarly, has its continuity and momentum equation, where the backreaction of drag on the gas enters the latter as an integral over stopping time:
\begin{align}
\partial_t \rho_{\rm g} + \nabla\cdot (\rho_{\rm g} {\bf v}_{\rm g})=&0,\label{eq:gascont}\\
\partial_t{\bf{v}_{\rm g}} + ({\bf{v}_{\rm g}}\cdot\nabla){\bf{v}_{\rm g}} =& -\frac{\nabla p}{{\rho_{\rm g}}} +\bm{\alpha}_{\rm g} + \frac{1}{{\rho_{\rm g}}}\int \sigma \frac{{\bf{u}} - {\bf{v}_{\rm g}}}{\taus}\rmd\taus\, ,
\label{eq:gasmom}
\end{align}
where ${\rho_{\rm g}}$ is the gas density, $p$ the pressure, and all additional accelerations on the gas are contained in $\bm{\alpha}_{\rm g}$. 

Our basic setup will be a standard unstratified shearing box \citep{1965MNRAS.130..125G,1995ApJ...440..742H,2017MNRAS.472.1432L}. Following \cite{2020MNRAS.497.2715K}, to account for vertical stratification, we add an extra vertical acceleration to the dust component $\bm {\alpha}_{\rmd} = z_0\Omega^2{\bf \hat z}$, where $z_0$ can be thought of as the height of the box above the mid plane\footnote{Strictly speaking, the corresponding height would be $-z_0$, to make sure the acceleration is in the direction of the mid plane. This particular choice is consistent with \cite{2020MNRAS.497.2715K}.}. For the gas, this vertical acceleration is balanced by a background pressure gradient, which we subtract from the total pressure. What is left is the backreaction on the gas, $\mu z_0\Omega^2$, so that the total acceleration of gas and dust are given by:
\begin{align}
\bm{\alpha}_{\rm g} =& 2\eta {\bf \hat x} -\mu z_0\Omega^2 {\bf \hat z}- 2\bm{\Omega}\times {\bf v}_{\rm g}-\nabla\Phi,\\
\bm{\alpha}_{\rm d} =& z_0\Omega^2 {\bf \hat z}- 2\bm{\Omega}\times {\bf u}-\nabla\Phi.
\end{align} 
The potential $\Phi = -S\Omega x^2$, where $S$ is the rate of orbital shear. The parameter $\eta$ is determined by the global radial pressure gradient, $\eta = -1/(2\rho_{\rm g})\rmd P/\rmd r$ \citep{2020MNRAS.499.4223P}. The equations governing the gas are then given by (\ref{eq:gascont}) and (\ref{eq:gasmom}), with accelerations as stated above (for a more formal derivation, see Appendix \ref{sec:appendix_eq}):
\begin{align}
\partial_t{\rho_{\rm g}} + \nabla\cdot({\rho_{\rm g}}{\bf{v}_{\rm g}}) =& 0\, ,\label{eq:gascont_sbox}\\
\partial_t {\bf{v}_{\rm g}}
+ ({\bf{v}_{\rm g}} \cdot \nabla){\bf{v}_{\rm g}}
=&
2\eta {\bf \hat x} -\mu z_0\Omega^2 {\bf \hat z} -\frac{\nabla p}{{\rho_{\rm g}}}- 2\bm{\Omega}\times {\bf{v}_{\rm g}}-\nabla\Phi \nonumber\\
&+ \frac{1}{{\rho_{\rm g}}}\int \sigma\frac{{\bf{u}}-{\bf{v}_{\rm g}}}{\taus}\rmd \taus.\label{eq:gasmom_sbox}
\end{align}
It should be noted that the pressure does not include the vertical stratification; stratification is only taken into account through the accelerations involving $z_0\Omega^2$. We take the equation of state to be isothermal, $p = c^2{\rho_{\rm g}}$, with constant sound speed $c$. We fix $c$ by requiring that $\eta/(c\Omega)=10^{-3/2}$, similar to the choice in \cite{2020MNRAS.497.2715K}\footnote{Since our parameter $\eta$ is dimensional by choice, with dimensions of acceleration, $\eta/(c\Omega)$ is equivalent to $\Pi$ in \cite{2020MNRAS.497.2715K}.}. The dust fluid equations can similarly be constructed from (\ref{eq:dustcont}) and (\ref{eq:dustmom}):
\begin{align}
\partial_t\sigma + \nabla\cdot(\sigma {\bf{u}} )=& 0,\label{eq:dustcont_sbox}\\
\partial_t {\bf{u}}
+ ({\bf{u}} \cdot \nabla){\bf{u}}
=&
z_0\Omega^2 {\bf \hat z} -2\bm{\Omega}\times {\bf{u}}-\nabla\Phi - \frac{{\bf{u}}-{\bf{v}_{\rm g}}}{\taus}.\label{eq:dustmom_sbox}
\end{align}
Apart from the extra vertical acceleration, these equations are the same as in \cite{2020MNRAS.499.4223P, 2021MNRAS.502.1579P}. With these shearing-box equations for the gas-dust system in hand, we can proceed to solve for the equilibrium state.

\subsection{Equilibrium state}

For the equilibrium state, we take ${\rho_{\rm g}}$ and $\sigma$ to be spatially constant. An equilibrium solution can then be found with horizontal velocities independent of $y$ and $z$ \citep{2020MNRAS.499.4223P}: 
\begin{align}
{v_{{\rm g}x}}
=&
\frac{2\eta}{\kappa}\frac{\mathcal{J}_1}{\left(1+ \mathcal{J}_0\right)^2 + \mathcal{J}_1^2},
\\
{v_{{\rm g}y}}
=&
-Sx -\frac{\eta}{\Omega}\frac{1 +  \mathcal{J}_0}{\left(1+ \mathcal{J}_0\right)^2 + \mathcal{J}_1^2},\\
u_x =& \frac{2\eta}{\kappa}
 \frac{\mathcal{J}_1 - \kappa\taus(1 +  \mathcal{J}_0)}{(1+\kappa^2\taus^2)(\left(1+ \mathcal{J}_0\right)^2 + \mathcal{J}_1^2)} ,\\
 u_y   =& -Sx
- \frac{\eta}{\Omega}\frac{1 +  \mathcal{J}_0 + \kappa\taus\mathcal{J}_1}{(1+\kappa^2\taus^2)(\left(1+ \mathcal{J}_0\right)^2 + \mathcal{J}_1^2)},
\end{align}
with integrals
\begin{align}
\mathcal{J}_m = \frac{1}{{\rho_{\rm g}}}\int \frac{\sigma (\kappa\taus)^{m}}{1+\kappa^2\taus^2} \rmd\taus.
\end{align}
Here $\kappa^2=2\Omega(S-\Omega)$ is the square of the epicyclic frequency, with shear rate $S$. In a Keplerian disc, $S=3\Omega/2$, in which case $\kappa=\Omega$. Vertical dust velocities are constant in space and given by
\begin{align}
u_z = \taus z_0\Omega^2,
\end{align}
while the vertical gas velocity is zero. The dust velocities corresponds to a uniform vertical settling flow for each dust size, at the appropriate terminal velocity in the uniform vertical gravitational field of the model.

\subsection{Linear perturbations}

Consider small perturbations ${\rho_{\rm g}} = {\rho_{\rm g}}^{(0)} + {\rho_{\rm g}}^{(1)}$ with ${\rho_{\rm g}}^{(1)} \ll {\rho_{\rm g}}^{(0)}$, and similar for other quantities. Ignoring quadratic terms in perturbed quantities, we find from the gas equations:
\begin{align}
\partial_t{\rho_{\rm g}^{(1)}}
&+ {\bf v}_{\rm g}^{(0)}\cdot \nabla{\rho_{\rm g}^{(1)}}
+ {\rho_{\rm g}^{(0)}}\nabla\cdot {\bf v}_{\rm g}^{(1)} = 0,\\
\partial_t {\bf v}_{\rm g}^{(1)}
&+ ({\bf v}_{\rm g}^{(0)}\cdot \nabla){\bf v}_{\rm g}^{(1)}
-S {v^{(1)}_{{\rm g}x}}{\bf \hat y}
=
-\frac{\nabla p^{(1)}}{{\rho_{\rm g}^{(0)}}}- 2\bm{\Omega}\times {\bf v}_{\rm g}^{(1)}\nonumber\\
&+ \frac{1}{{\rho_{\rm g}^{(0)}}}\int \sigma^{(1)}\frac{\Delta{\bf{u}}^{(0)}}{\taus}\rmd\taus
+ \frac{1}{{\rho_{\rm g}^{(0)}}}\int \sigma^{(0)}\frac{{\bf{u}}^{(1)}-{\bf v}_{\rm g}^{(1)}}{\taus}\rmd\taus,
\end{align}
with equilibrium relative velocity $\Delta {\bf{u}}^{(0)} = {\bf{u}}^{(0)}-{\bf v}_{\rm g}^{(0)}$. Dust perturbations are governed by:
\begin{align}
\partial_t\sigma^{(1)}
&+ {\bf{u}}^{(0)}\cdot\nabla\sigma^{(1)}
+ \sigma^{(0)}\nabla\cdot {\bf{u}}^{(1)} = 0,
\label{eq:linear_sigma}\\
\partial_t {\bf{u}}^{(1)}
&+ ({\bf{u}}^{(0)}\cdot\nabla){\bf{u}}^{(1)}
 - Su_{x}^{(1)}{\bf \hat y}
=\nonumber\\
&-2\bm{\Omega}\times {\bf{u}}^{(1)}
- \frac{{\bf{u}}^{(1)}-{\bf v}_{\rm g}^{(1)}}{\taus}
- \frac{{\rho_{\rm g}^{(1)}}}{{\rho_{\rm g}^{(0)}}} \frac{\Delta {\bf{u}}^{(0)}}{\taus}.
\end{align}
Consider perturbations of the form $X^{(1)}({\bf x},t,\taus) = \hat X(\taus) \exp(\rmi {\bf k}\cdot {\bf x} - \rmi \omega t)$, with wavenumber ${\bf k} = (k_x, k_y, k_z)^T$ and frequency $\omega$. We consider only perturbations with $k_y=0$. Using the above form of the perturbations, the dust and gas perturbation equations transform to:
\begin{align}
k_x {v^{(0)}_{{\rm g}x}} \frac{{\hat{\rho}_{\rm g}}}{{\rho_{\rm g}^{(0)}}}
&+ {\bf k}\cdot {\bf{\hat{v}}_{\rm g}} =
\omega \frac{{\hat{\rho}_{\rm g}}}{{\rho_{\rm g}^{(0)}}},\label{eq:eig_gasdens}\\
k_x {v^{(0)}_{{\rm g}x}} {\bf{\hat{v}}_{\rm g}}
&+ \rmi S {\hat{v}_{{\rm g}x}}{\bf \hat y}
+\frac{{\bf k} c^2{\hat{\rho}_{\rm g}}}{{\rho_{\rm g}^{(0)}}}
- 2\rmi \bm{\Omega}\times {\bf{\hat{v}}_{\rm g}} + {\bf \hat b} =
\omega {\bf{\hat{v}}_{\rm g}},\label{eq:eig_gasvel}\\
 {\bf k}\cdot {\bf u}^{(0)} \hat \sigma
 &+ \sigma^{(0)}{\bf k}\cdot {\bf \hat u} =
\omega \hat\sigma,\label{eq:eig_dustdens}\\
({\bf k}\cdot {\bf u}^{(0)})  {\bf \hat u}
&+ \rmi S\hat u_{x}{\bf \hat y}
-2\rmi\bm{\Omega}\times {\bf \hat u} \nonumber\\
&-\rmi \frac{{\bf \hat u}-{\bf{\hat{v}}_{\rm g}}}{\taus}
-\rmi \frac{{\hat{\rho}_{\rm g}}}{{\rho_{\rm g}^{(0)}}} \frac{\Delta {\bf u}^{(0)}}{\taus}
=\omega {\bf \hat u}.\label{eq:eig_dustvel}
\end{align}
with the backreaction integral
\begin{align}
{\bf \hat b}=\frac{\rmi}{{\rho_{\rm g}^{(0)}}}\int \frac{\sigma^{(0)}}{\taus}\left[\frac{\hat \sigma}{\sigma^{(0)}}\Delta{\bf u}^{(0)}+ {\bf \hat u}-{\bf{\hat{v}}_{\rm g}}\right]\rmd \taus.
\end{align}
By taking the complex conjugate of each term in (\ref{eq:eig_gasdens})-(\ref{eq:eig_dustvel}), it is straightforward to see that a solution with eigenvalue $\omega$ and eigenvector $(\hat\rho_{\rm g}, {\bf \hat v}_{\rm g}, \hat \sigma, {\bf \hat u})$ can be transformed into a solution with eigenvalue $-\omega^*$ and eigenvector $(\hat\rho_{\rm g}^*, {\bf \hat v}_{\rm g}^*, \hat \sigma^*, {\bf \hat u}^*)$, where an asterisk denotes complex conjugate. Since $\omega$ and $-\omega^*$ have the same imaginary part and therefore growth rate, we follow the convention of e.g.\cite{2005ApJ...620..459Y, 2018MNRAS.477.5011S, 2019ApJ...878L..30K} and focus exclusively on $k_x,k_z>0$.

With the help of the dust continuity and momentum equations (see Appendix \ref{sec:appendix_disp}), the backreaction integral can be written as
\begin{align}
{\bf \hat b} = \mathsf{M} {\bf{\hat{v}}_{\rm g}},
\end{align}
with matrix $\mathsf{M}$ defined in (\ref{eq:gasmomM}). In the limit of $\mu \ll1$, we therefore have a perturbed eigenvalue problem for the gas:
\begin{align}
    \left(\begin{array}{cccc}
    k_x v_{{\rm g},x}^{(0)} & k_x & 0 & k_z \\
    k_x c^2 & k_x v_{{\rm g},x}^{(0)} & 2\rmi \Omega & 0\\
    0 & (S-2\Omega)\rmi & k_x v_{{\rm g},x}^{(0)} & 0\\
    k_z c^2 & 0 & 0 & k_x v_{{\rm g},x}^{(0)} \end{array}\right)
    \left(\begin{array}{c} {\hat{\rho}_{\rm g}}/\rho_{\rm g}^{(0)} \\ {\bf{\hat{v}}_{\rm g}}\end{array}\right)\nonumber\\
    + \left(\begin{array}{cc}
    0 & 0\\
    0 & \mathsf{M} \end{array}\right)\left(\begin{array}{c} {\hat{\rho}_{\rm g}}/\rho_{\rm g}^{(0)} \\ {\bf{\hat{v}}_{\rm g}}\end{array}\right)=\omega \left(\begin{array}{c} {\hat{\rho}_{\rm g}}/\rho_{\rm g}^{(0)} \\ {\bf{\hat{v}}_{\rm g}}\end{array}\right),
    \label{eq:SI_eig_pert}
\end{align}
where the perturbation matrix contains the backreaction on the gas and is therefore $\propto \mu$. In regions of parameter space where the resonance is not important, ordinary eigenvalue perturbation theory can be used to approximate the growth rates for $\mu \ll 1$. Numerical results on growth rates are obtained by the package {\sc psitools} \citep{2021MNRAS.502.1579P}.

It is often useful to define a length scale $\eta/\Omega^2$ and a time scale $\Omega^{-1}$ and work with dimensionless wave numbers ${\bf K}  = {\bf k}\eta/\Omega^2$ and dimensionless stopping time, or Stokes number, $\mathrm{St}=\Omega\taus$. It is worth noting that the non-dimensional 'height' above the midplane, $\tilde z_0 = z_0\Omega^2/\eta$ is a large number if we take $z_0 \sim H$, where $H$ is the vertical scale height of the gas disc.  

\subsection{Eigenvalue perturbation theory for the inertial wave RDI}
\label{sec:app_pert}

The perturbed eigenvalue problem for the inertial wave RDI (\ref{eq:SI_eig_pert}) can be simplified by taking the gas to be incompressible. In this case, the gas equations (\ref{eq:eig_gasdens})-(\ref{eq:eig_gasvel}) can be simplified by taking the crossproduct of the wave vector with the gas momentum equation:
\begin{align}
k_x {v^{(0)}_{{\rm g}x}} {\bf{\hat{w}}_{\rm g}}
&+ \rmi S {\hat{v}_{{\rm g}x}} ({\bf k}\times{\bf \hat y})
- 2\rmi {\bf k}\times(\bm{\Omega}\times {\bf{\hat{v}}_{\rm g}})
+{\bf k}\times {\bf \hat b}
=
\omega {\bf{\hat{w}}_{\rm g}},
\end{align}
with vorticity perturbation ${\bf{\hat{w}}_{\rm g}}={\bf k}\times {\bf{\hat{v}}_{\rm g}}=(-k_z{\hat{v}_{{\rm g}y}}, k_z{\hat{v}_{{\rm g}x}} - k_x{\hat{v}_{{\rm g}z}}, k_x{\hat{v}_{{\rm g}y}})^T$. Expand the components of the Coriolis terms to find:
\begin{align}
k_x {v^{(0)}_{{\rm g}x}} {\bf{\hat{w}}_{\rm g}}
+ \rmi \left(\begin{array}{c} (2\Omega-S) k_z{\hat{v}_{{\rm g}x}} \\ 2\Omega k_z{\hat{v}_{{\rm g}y}} \\ -(2\Omega-S) k_x{\hat{v}_{{\rm g}x}} \end{array}\right)
+{\bf k}\times {\bf \hat b}
=
\omega {\bf{\hat{w}}_{\rm g}}.
\end{align}
Noting that the last component of ${\bf{\hat{w}}_{\rm g}}$ is redundant, so that we only need to keep the first two components, and that ${\hat{v}_{{\rm g}x}} = k_zw_{{\rm g}y}/(k_x^2+k_z^2)$:
\begin{align}
k_x {v^{(0)}_{{\rm g}x}} {\bf{\hat{w}}_{\rm g}}
+ \rmi \left(\begin{array}{c} (2\Omega-S) \hat k_z^2w_{{\rm g}y} \\ -2\Omega w_{{\rm g}x}\end{array}\right)
+\mathsf{K}{\bf \hat b}
=
\omega {\bf{\hat{w}}_{\rm g}},
\label{eq:disp_inc}
\end{align}
with $\hat k_z = k_z/|{\bf k}|$ and matrix
\begin{align}
    \mathsf{K}=\left(\begin{array}{ccc}
0 & -k_z & 0\\
k_z & 0 & -k_x \end{array}\right),
\end{align}
representing the crossproduct with the wave vector. We can write (\ref{eq:disp_inc}) in simple matrix form:
\begin{align}
\left(\begin{array}{cc} k_x {v^{(0)}_{{\rm g}x}} & \rmi(2\Omega-S) \hat k_z^2 \\ -2\rmi\Omega & k_x {v^{(0)}_{{\rm g}x}}\end{array}\right){\bf{\hat{w}}_{\rm g}}
+\mathsf{K}\mathsf{M}\mathsf{K}'{\bf{\hat{w}}_{\rm g}}
=
\omega {\bf{\hat{w}}_{\rm g}},
\end{align}
with matrix 
\begin{align}
\mathsf{K}' = \left(\begin{array}{cc}
0 & k_z/k^2\\
-1/k_z & 0\\
0 & -k_x/k^2\end{array}\right).
\end{align}
We now have a perturbed eigenvalue problem $(\mathsf{T}_0 + \mathsf{T}_1){\bf{\hat{w}}_{\rm g}} = \omega{\bf{\hat{w}}_{\rm g}}$. The unperturbed matrix $\mathsf{T}_0$ gives the familiar inertial waves, advected with the background gas flow. The corresponding eigenvalues are
\begin{align}
    \omega_0=k_x {v^{(0)}_{{\rm g}x}} \pm \hat k_z \kappa,
    \label{eq:omega0}
\end{align}
and the left and right eigenvectors
\begin{align}
{\bf e}_L =& \left(\begin{array}{cc}\mp \frac{\rmi}{\hat k_z} & \frac{\kappa}{2\Omega}\end{array}\right),\\
    {\bf e}_R =& \left(\begin{array}{c} \pm \rmi \hat k_z \frac{\kappa}{2\Omega} \\ 1\end{array}\right).
\end{align}

The first order correction to the eigenvalue due to $\mathsf{T}_1$ is simply given by
\begin{align}
\omega_1 = \frac{{\bf e}_L^* \mathsf{T}_1 {\bf e}_R}{{\bf e}_L^* {\bf e}_R},
\end{align}
where ${\bf e}_L$ and ${\bf e}_R$ are the left and right eigenvector of $\mathsf{T}_0$ corresponding to the unperturbed eigenvalue $\omega_0$ given above, and $^*$ denotes complex conjugate. Together with the perturbation matrix
\begin{align}
  \mathsf{T}_1 = \mathsf{K}\mathsf{M}\mathsf{K}',  
\end{align}
we can calculate the correction to the eigenvalues in the limit of $\mu \ll 1$ (as long as the perturbation is regular).

Since the eigenvalue perturbation is linear in the elements of $\mathsf{T}_1$, and therefore linear in the elements of $\mathsf{M}$, it is straightforward to distill the contributions of the dust density perturbation and the velocity perturbations to $\omega_1$ (see Appendix \ref{sec:appendix_disp} for the definitions of matrices $\mathsf{D}$ and $\mathsf{A}$):
\begin{align}
\mathsf{M}_\sigma =& \int \frac{\rmi\sigma^{(0)}}{{\rho_{\rm g}}^{(0)}\taus}\mathsf{V}_\sigma\mathsf{A}^{-1}\mathsf{D}\rmd\taus,\\
\mathsf{M}_u =& \int \frac{\rmi\sigma^{(0)}}{{\rho_{\rm g}}^{(0)}\taus}\mathsf{A}^{-1}\mathsf{D}\rmd\taus,\\
\mathsf{M}_v =& -\int \frac{\rmi\sigma^{(0)}}{{\rho_{\rm g}}^{(0)}\taus}\rmd\taus \mathsf{I},
\end{align}
with
\begin{align}
    \mathsf{V}_\sigma = \frac{1}{\omega-k_xu_x^{(0)} -k_z u_z^{(0)}}\left(\begin{array}{ccc}
\Delta u_x^{(0)} k_x & 0 & \Delta u_x^{(0)} k_z\\
\Delta u_y^{(0)} k_x & 0 & \Delta u_y^{(0)} k_z\\
\Delta u_z^{(0)} k_x & 0 & \Delta u_z^{(0)} k_z
\end{array}\right).
\end{align}
If we use $\mathsf{M}_\sigma$ to calculate $\mathsf{T}_1$ and $\omega_1$, we get the contribution from only the dust density perturbation. It is convenient to lump the dust and gas velocity perturbations together and look at the contribution of the relative velocity perturbation governed by $\mathsf{M}_u+\mathsf{M}_v$. 

We can try and isolate the contribution of the resonance to the growth rate by focusing on the dust density perturbation:
\begin{align}
\omega_{1,\sigma} = \frac{{\bf e}_L^* \mathsf{K}\mathsf{M}_\sigma \mathsf{K}'{\bf e}_R}{{\bf e}_L^* {\bf e}_R} = \int \frac{\rmi\sigma^{(0)}}{{\rho_{\rm g}}^{(0)}\taus}\frac{{\bf e}_L^* \mathsf{K}\mathsf{V}_\sigma\mathsf{A}^{-1}\mathsf{D}\mathsf{K}'{\bf e}_R}{{\bf e}_L^* {\bf e}_R}\rmd \taus.
\end{align}
Isolate the resonance contribution by defining matrix $\mathsf{R}$ such that $\mathsf{V}_\sigma = \mathsf{R}/(\omega-k_xu_x^{(0)} - k_z u_z^{(0)})$:
\begin{align}
\omega_{1,\sigma} = \int \frac{g(\taus)}{\omega_0-k_xu_x^{(0)} - k_z u_z^{(0)}}\rmd \taus,
\end{align}
with
\begin{align}
g(\taus) = \frac{\rmi\sigma^{(0)}}{{\rho_{\rm g}}^{(0)}\taus}\frac{{\bf e}_L^* \mathsf{K}\mathsf{R}\mathsf{A}^{-1}\mathsf{D}\mathsf{K}'{\bf e}_R}{{\bf e}_L^* {\bf e}_R}, 
\end{align}
a complex scalar function. Approximate $\Delta u_x = -2\eta \taus$, and use $u_z=\Omega^2z_0 \taus$, as well as the expression (\ref{eq:omega0}) for $\omega_0$:
\begin{align}
\frac{\omega_{1,\sigma}}{\Omega} =\int \frac{\tilde g(\stokes)}{\pm \hat k_z + (2K_x - K_z\tilde z_0)\stokes}\rmd \stokes,
\end{align}
with non-dimensional $\tilde g = g/\Omega^2$. Note that, for a resonance to occur, we need $\pm \hat k_z/(K_z\tilde z_0-2K_x) > 0$. This means that, at a given Stokes number, the two branches of inertial waves give rise to resonances at different locations in $K$-space. These two resonant branches come together at $K_z\tilde z_0 = 2K_x$, which is the location of the double resonance \citep{2018MNRAS.477.5011S}, which we do not consider here, since ordinary perturbation theory does not hold there. On the other hand, for a fixed wave vector ${\bf K}$, if $\pm \hat k_z/(K_z\tilde z_0-2K_x) > 0$, there is a Stokes number that gives rise to a resonance, at least if it is inside the original size range.

It is convenient to substitute $\xi=(2K_x-K_z\tilde z_0)\stokes \pm \hat k_z$, so that at the resonant Stokes number $\stokes_{\rm res}$ we have $\xi=0$. If we further assume that $\tilde g$ can be expanded in a Taylor series around $\stokes=\stokes_{\rm res}$, or, equivalently, around $\xi=0$:
\begin{align}
    \tilde g(\stokes)= \sum_{n=0}^\infty \tilde g_n \left(\stokes-\stokes_{\rm res}\right)^n = \sum_{n=0}^\infty \frac{\tilde g_n}{\left(2K_x-K_z\tilde z_0\right)^n} \xi^n,
\end{align}
we obtain
\begin{align}
\frac{\omega_{1,\sigma}}{\Omega} =\sum_{n=0}^\infty \frac{\tilde g_n}{\left(2K_x-K_z\tilde z_0\right)^{n+1}}\int \frac{\xi^n}{\xi}\rmd \xi,
\end{align}
where we have explicitly kept the denominator in place to highlight the resonance. For $n>0$ the integral is straightforward:
\begin{align}
\frac{\omega_{1,\sigma}}{\Omega} =\frac{\tilde g_0}{2K_x-K_z\tilde z_0}\int \frac{\rmd\xi}{\xi} + \sum_{n=1}^\infty \frac{\tilde g_n}{(2K_x-K_z\tilde z_0)^{n+1}}\frac{\xi_{\rm max}^n -\xi_{\rm min}^n}{n}.
\label{eq:integrand}
\end{align}
We solve the remaining integral adopting the Landau prescription, well-known from plasma physics \citep{Landau:1946jc}, and also used in the theory of collisionless stellar systems \citep{2008gady.book.....B}, and disc-planet interaction \citep{2008A&A...485..877P}. Including a small damping term $-\sigma^{(1)}/t_{\rm damp}$ on the right hand side of (\ref{eq:linear_sigma}) leads to a small imaginary part $\rmi \epsilon =\rmi/(\Omega t_{\rm damp})$ in the denominator of the integrand in (\ref{eq:integrand}). To arrive at the relevant limit of no damping, we let $\epsilon\rightarrow 0$:
\begin{align}
    \int_{\xi_{\rm min}}^{\xi_{\rm max}} \frac{\rmd\xi}{\xi+\rmi \epsilon}=&
    \int_{\xi_{\rm min}}^{\xi_{\rm max}} \frac{\xi-\rmi\epsilon}{\xi^2+\epsilon^2}\rmd\xi \nonumber\\
    =& \left[\frac{\log(\xi^2+\epsilon^2)}{2}-\rmi \tan^{-1}\left(\frac{\xi}{\epsilon}\right)\right]_{\xi_{\rm min}}^{\xi_{\rm max}}.
\end{align}
Note that $\xi_{\rm max}$ corresponds to the value of $\xi$ at $\stokes=\stokes_{\rm max}$. Depending on the sign of $2K_x-K_z\tilde z_0$, $\xi_{\rm max}$ may be positive or negative. Therefore, in the limit $\epsilon\rightarrow 0$, we get that
\begin{align}
    \int_{\xi_{\rm min}}^{\xi_{\rm max}} \frac{\rmd\xi}{\xi+\rmi \epsilon}=\log\left(\frac{|\xi_{\rm max}|}{|\xi_{\rm min}|}\right)
    -\rmi\pi \left(\frac{\xi_{\rm max}}{|\xi_{\rm max}|} - \frac{\xi_{\rm min}}{|\xi_{\rm min}|}\right),
\end{align}
where the last factor in parenthesis makes sure that we only get a contribution if the resonance is in the domain, i.e. when $\xi_{\rm max}$ and $\xi_{\rm min}$ have opposite sign, and that the sign gets swapped if $\xi_{\rm max}$ and $\xi_{\rm min}$ both change sign. In terms of Stokes numbers we get
\begin{align}
\frac{\omega_{1,\sigma}}{\Omega} =&\frac{\tilde g_0}{2K_x-K_z\tilde z_0}\log\left(\frac{|\stokes_{\rm max}-\stokes_{\rm res}|}{|\stokes_{\rm min}-\stokes_{\rm res}|}\right)\nonumber\\
-&\frac{\tilde g_0}{|2K_x-K_z\tilde z_0|}\frac{\rmi\pi}{2} \left[\frac{\stokes_{\rm max}-\stokes_{\rm res}}{|\stokes_{\rm max}-\stokes_{\rm res}|}+\frac{\stokes_{\rm res}-\stokes_{\rm min}}{|\stokes_{\rm res}-\stokes_{\rm min}|}\right]\nonumber\\
+& \sum_{n=1}^\infty \frac{\tilde g_n}{2K_x-K_z\tilde z_0}\frac{(\stokes_{\rm max}-\stokes_{\rm res})^n-(\stokes_{\rm min}-\stokes_{\rm res})^n}{n},
\label{eq:size_dens_cont}
\end{align}
where the factor in square brackets makes sure the resonance only contributes if it is inside the domain, i.e. if $\stokes_{\rm min} < \stokes_{\rm res} < \stokes_{\rm max}$.

In order to isolate the contribution from the resonance, we can approximate $\tilde g_0$ up to first order in $\stokes$ and evaluate it at $\stokes=\stokes_{\rm res}$:
\begin{align}
    \tilde g_0 = 
    \mp \frac{\sigma^{(0)}(\stokes_{\rm res})}{{\rho_{\rm g}^{(0)}}}K_x\stokes_{\rm res}\left(\hat k_z +\tilde z_0\hat k_x/2\right), 
\end{align}
so that the contribution of the resonance to the growth rate is simply
\begin{align}
\frac{\omega_{1,{\rm res}}}{\Omega} =&
\pm\rmi\pi\frac{\sigma^{(0)}(\stokes_{\rm res})}{{\rho_{\rm g}^{(0)}}}\frac{K_x\stokes_{\rm res}\left(\hat k_z +\tilde z_0\hat k_x/2\right)}{|2K_x-K_z\tilde z_0|}\nonumber\\
=& \pm\rmi\pi\frac{\sigma^{(0)}(\stokes_{\rm res})}{{\rho_{\rm g}^{(0)}}}\frac{K_x \hat k_z\left(\hat k_z +\tilde z_0\hat k_x/2\right)}{(2K_x-K_z\tilde z_0)^2},
\label{eq:res_cont}
\end{align}
where for clarity we have omitted the factor determining if the resonance is inside the domain. This is the contribution of the resonance to the growth rate. The full perturbed eigenvalue includes in addition contributions from the non-resonant size density perturbation (see (\ref{eq:size_dens_cont})) as well as relative velocity perturbations, the latter of which always have a stabilizing effect.  

Summarizing, if we assume regular eigenvalue perturbation theory applies for $\mu\ll 1$, under the further assumption that $\stokes \ll 1$, an inertial wave instability can develop for a constant size distribution if the imaginary part of (\ref{eq:res_cont}) is positive, and in addition dominates over the stabilizing effects of the relative velocity perturbation. Note that the two possible signs of (\ref{eq:res_cont}) correspond to the two branches of inertial waves with positive/negative frequency.

\section{Streaming Instability}
\label{sec:stream}

We first fix $z_0=0$ to disable settling, and look at the classic SI. We stay firmly in the RDI regime $\mu \ll 1$, which, although perhaps not a case usually considered for the SI \citep[but see][]{2024MNRAS.tmp.1938M}, can provide additional insight into its behaviour for a distribution of particle sizes.

Numerical growth rates are computed using the package {\sc psitools} \citep{2021MNRAS.502.1579P}. The presence of the resonance makes computation by standard eigenvalue solvers problematic \citep{2019ApJ...878L..30K, 2021MNRAS.502.1579P}, and {\sc psitools} was specifically designed to handle this. It has been tested and verified against direct eigenvalue solvers where the latter can be applied \citep{2021MNRAS.502.1579P}.

\subsection{Monodisperse limit}

\begin{figure}
  \resizebox{\hsize}{!}{\includegraphics{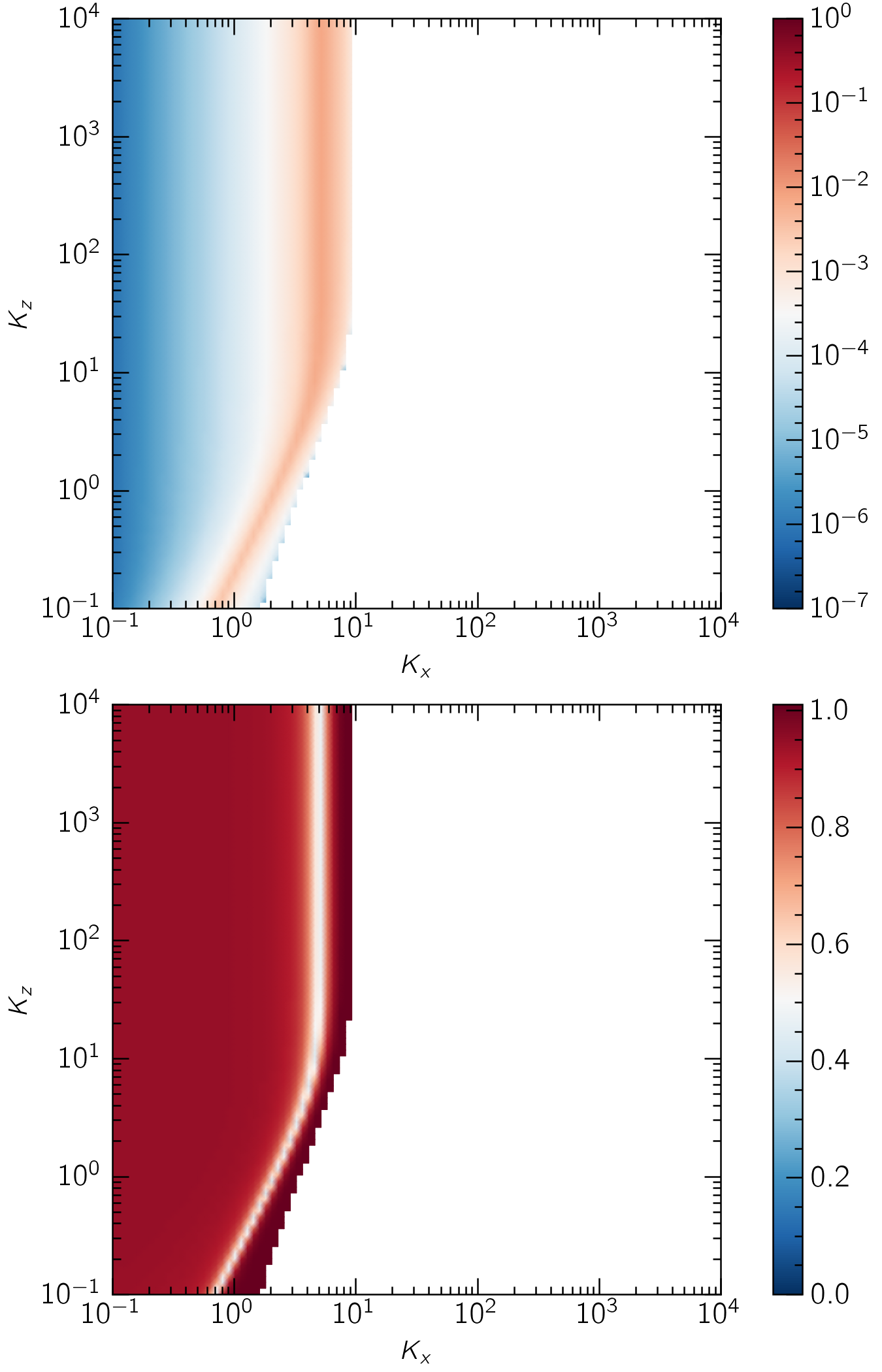}}
  \caption{Top panel: growth rates $\Im(\omega)/\Omega$ for the monodisperse streaming instability with $\stokes=0.1$ and $\mu=0.01$. Bottom panel: $\rmd\log\Im(\omega)/\rmd\log\mu$ for $\stokes=0.1$ at $\mu=0.01$.}
  \label{fig:SI_monoresonance}
\end{figure}

In the top panel of Figure \ref{fig:SI_monoresonance}, we show the growth rates for the monodisperse case with $\stokes=0.1$ and $\mu=0.01$. Maximum growth rates are obtained for the resonant condition $\hat k_z\Omega = -k_x\Delta u_x^{(0)}$ (note that $\Delta u_x^{(0)} < 0$), in which case $\Im(\tilde\omega) \sim \sqrt{\mu}\stokes$ \citep{2018MNRAS.477.5011S}. The resonance can be seen even more clearly in the bottom panel of Figure \ref{fig:SI_monoresonance}, where we plot $\rmd\log\Im(\omega)/\rmd\log\mu$. This quantity was calculated numerically by central differences. Away from the resonance, since $\mu \ll 1$, we expect perturbations to the eigenvalue to be $\propto \mu$ (red regions), while at the resonance the perturbation follows $\propto \sqrt{\mu}$ (white region). To the left of the resonance, growth is dominated by the secular mode \citep{2005ApJ...620..459Y}, which is a perturbation to the dust advection mode. To the right of the resonance, a small non-resonant band of growth exists, which is formed by a regular perturbation of the same inertial mode that gives rise to the resonance. The largest growth rates are found at the location of the resonance, with $\max(\Im(\tilde\omega)) \approx 10^{-2} = \sqrt{\mu}\stokes$. 

\begin{figure}
  \resizebox{\hsize}{!}{\includegraphics{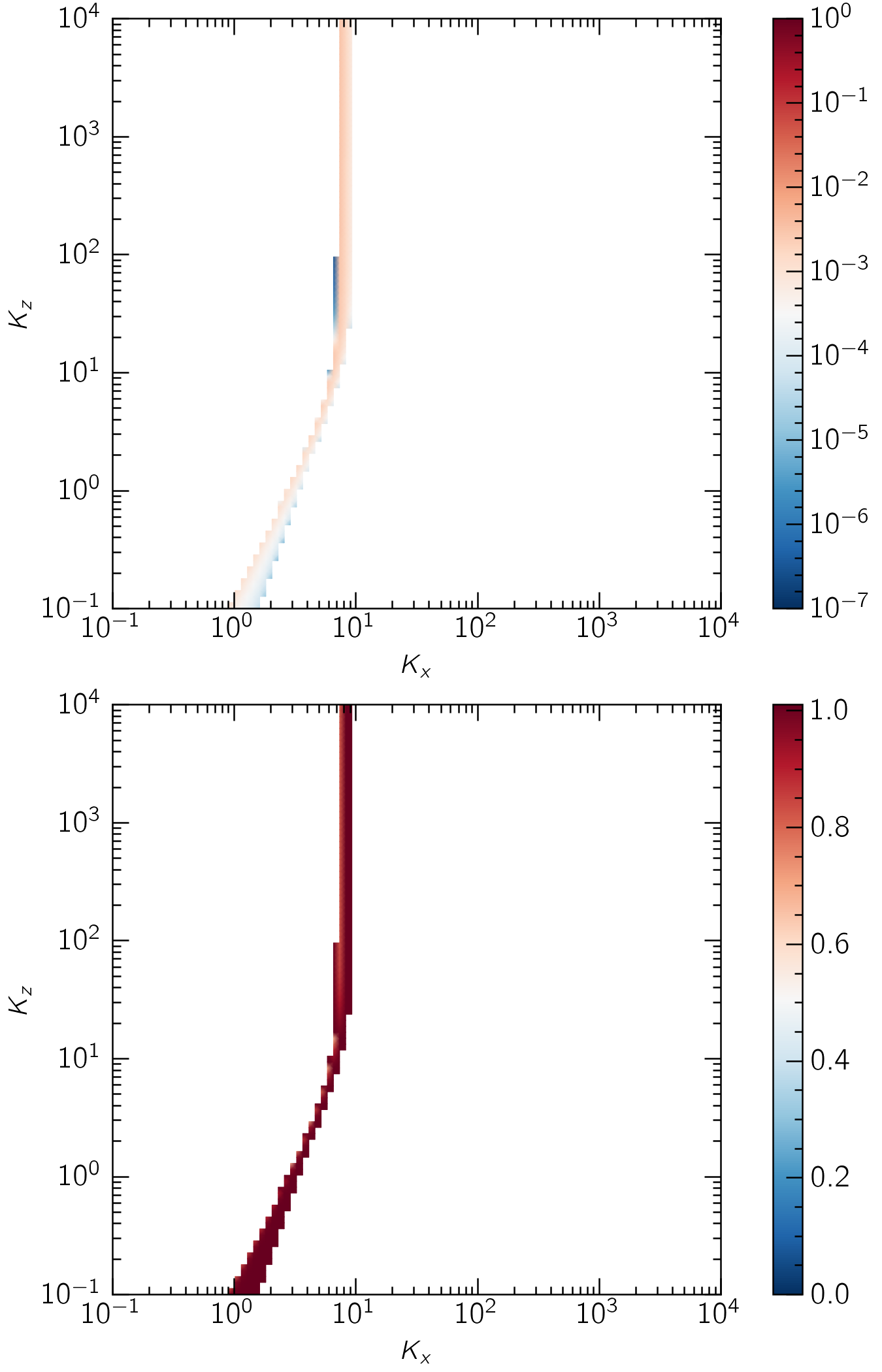}}
  \caption{Top panel: growth rates for the polydisperse streaming instability for $\mu=0.01$, $\stokes_0=0.1$ and $\Delta=0.3$. Bottom panel: $\rmd\log\Im(\omega)/\rmd\log\mu$ for $\stokes_0=0.1$, $\Delta=0.3$ at $\mu=0.01$.}
  \label{fig:SI_resonance}
\end{figure}

\subsection{Polydisperse streaming instability}

We now turn to a simple constant size distribution:
\begin{align}
  \sigma^{(0)}\left(\stokes\right) = \mu\rho_{\rm g}^{(0)}\left\{\begin{array}{ll}
                       \frac{1}{\stokes_{\rm max}-\stokes_{\rm min}} & \stokes_{\rm min} < \stokes < \stokes_{\rm max} \\
                       0 & {\rm otherwise.}
                     \end{array}\right.
                     \label{eq:sizedistSI}
\end{align}
As in Paper I, we parametrize the minimum and maximum stopping time as $\stokes_{0}(1\pm\Delta)$, where $\Delta$ is a measure of the width of the size distribution.

In Figure \ref{fig:SI_resonance}, we show the growth rates for $\stokes_0=0.1$ and $\mu=0.01$ and $\Delta=0.3$. Note $\Delta=0$ would correspond to the monodisperse result shown in Figure \ref{fig:SI_monoresonance}. Note also that this is a size distribution that is much narrower than expected in protoplanetary discs: if the resonant size is 1 mm, $\Delta=0.3$ means that the minimum size is $0.7$ mm and the maximum size is $1.3$ mm. In reality, we would expect a distribution that spans many decades in size \citep[see e.g.][]{2021MNRAS.502.1469M}. The top panel of Figure \ref{fig:SI_resonance} reveals that growing modes for small $K_x$ have disappeared. This is first of all due to the fact that most of this region is dominated by the secular mode in the monodisperse case, for which no equivalent exists for polydisperse dust \citep[see][as well as Paper I]{2020MNRAS.499.4223P}. Careful inspection of the wavenumbers that give growth, reveals that we have also lost the resonant region compared to the monodisperse case. While the bottom panel of Figure \ref{fig:SI_resonance} shows that growth is now $\propto \mu$ everywhere, this is not due to the fact that integrating over the resonance yields a regular perturbation: integrating over the resonance actually stabilizes the resonant region, which we discuss below. What is left is the non-resonant region to the right of the resonance in Figure \ref{fig:SI_monoresonance}, with growth rates that are comparable to the corresponding monodisperse case. 

The contribution of the resonance to the growth rate is given by (\ref{eq:res_cont}), which reads for $z_0=0$:
\begin{align}
\frac{\omega_{1,{\rm res}}}{\Omega} =
-\rmi \frac{\mu \pi}{8\stokes_0\Delta}\frac{\hat k_z^2}{K_x},
\label{eq:si_res_cont}
\end{align}
where we have also picked the inertial wave with $\omega_0=-\hat k_z\Omega$, which is the only one capable of forming a resonance for $\tilde z_0=0$, and is the one that goes unstable for the SI. In theory, the inertial wave with positive frequency could lead to non-resonant growth, but this is found not to be the case. From (\ref{eq:si_res_cont}), it is clear that the resonance has a stabilizing effect on the system for all wave numbers and Stokes numbers, as long as the size distribution is wide enough so that the perturbation is regular (note that (\ref{eq:si_res_cont}) diverges for $\Delta \rightarrow 0$, which takes us into the domain of the monodisperse SI). 

If growth rates are $\propto\mu$, which means we are either away from resonance, or the size distribution is wide enough so that the contribution of the resonance is regularised, we can use regular eigenvalue perturbation theory to obtain perturbed eigenvalues, see section \ref{sec:app_pert}. Furthermore, we can then separate the various contributions to the growth rate: dust density perturbation, and the contribution from the relative velocity. As it turns out, the relative velocity perturbation always gives a stabilizing contribution, which means that in order to get growth, the contribution of the dust density perturbation needs to be strong enough.

\begin{figure}
  \resizebox{\hsize}{!}{\includegraphics{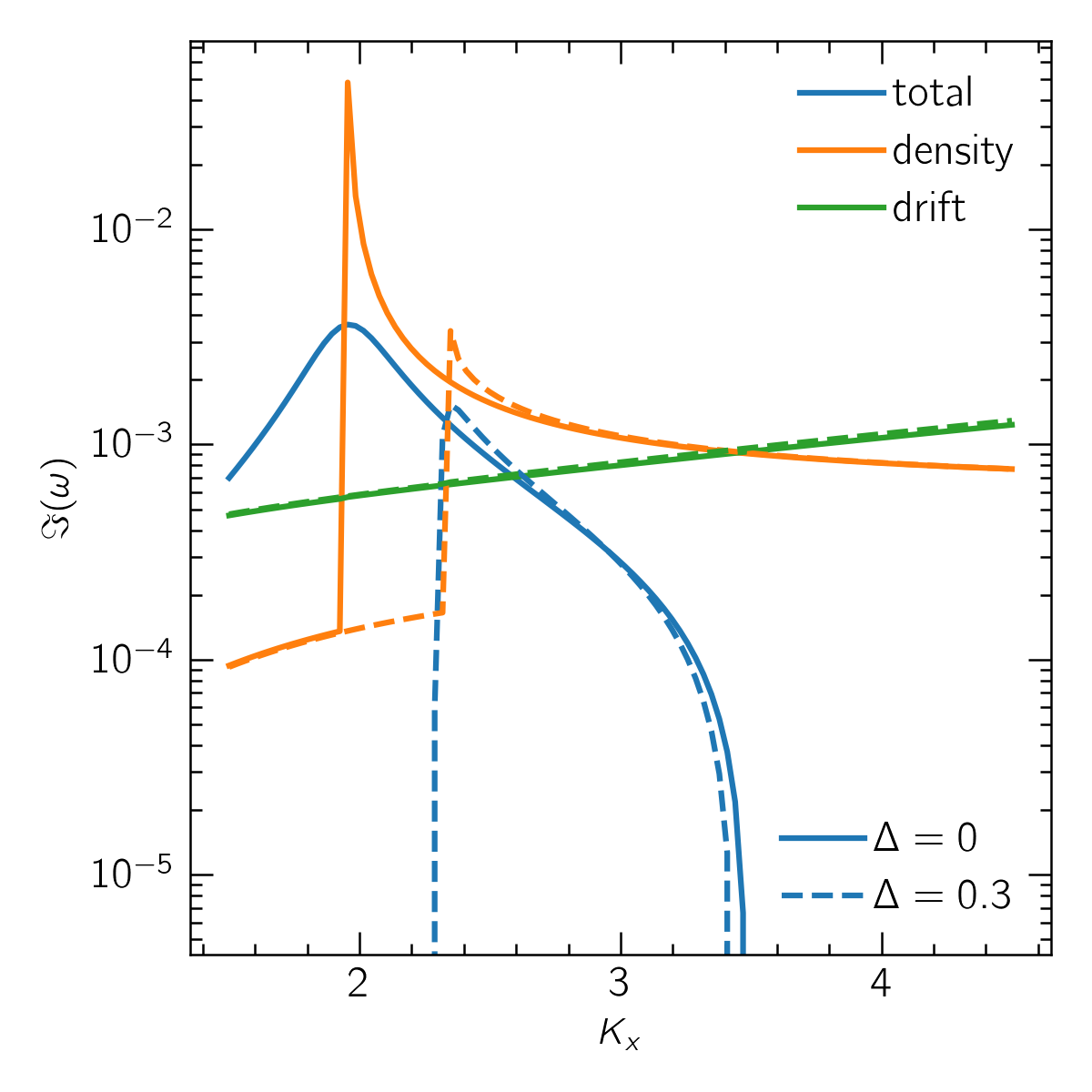}}
  \caption{Growth rate, in units of $\Omega^{-1}$, at $K_z=0.8$, for the monodisperse case of Figure \ref{fig:SI_monoresonance} (solid curves) and the polydisperse case of Figure \ref{fig:SI_resonance} (dashed curves). Yellow curves: contribution of the dust density perturbation to the growth rate. Green curves: contribution of the relative velocity perturbation to the growth rate (absolute contribution, since it is a stabilizing effect). Blue curves: total growth rate.}
  \label{fig:SI_contribution}
\end{figure}

This is illustrated in Figure \ref{fig:SI_contribution}, which considers the same two cases as in Figures \ref{fig:SI_monoresonance} and \ref{fig:SI_resonance}, for fixed $K_z=0.8$. The solid and dashed curves correspond to the monodisperse and polydisperse cases, respectively. The total growth rates (blue curves) were calculated exactly, while the contributions from the dust density (orange curves) and drift velocity (green curves) were calculated using perturbation theory (see section \ref{sec:app_pert}). Note that the drift velocity contribution is always negative, i.e. promoting stability rather than growth, which means we use the absolute value in Figure \ref{fig:SI_contribution}. The monodisperse growth rate peaks at the resonance, the location of which can also be inferred from the asymptote in the dust density perturbation (solid orange curve). Just to the left of the resonance, the dust density contribution is due to the inertial mode with opposite sign, which does not have a resonance. In fact, if we stick to the same mode, the dust density perturbation is an odd function around the resonance, as can be easily seen from the dust continuity equation. It should be noted however that close to the resonance, regular perturbation theory becomes invalid. In the full problem, to the left of the resonance (i.e. smaller $K_x$), the dust advection mode takes over, while for large enough $K_x$ growth is no longer possible, as the dust density contribution (solid orange curve) crosses the stabilizing contribution of the drift velocity (solid green curve). It is worth noting that this stabilizing effect is not present in the terminal velocity approximation, where the drift velocity is determined by the pressure gradient only, and as a result the terminal velocity approximation shows growth up to much larger wave numbers \citep{2017ApJ...849..129L, 2020MNRAS.499.4223P}. 

The polydisperse case with $\Delta=0.3$ follows the monodisperse results closely for the drift perturbation (green solid and dashed curves in Figure \ref{fig:SI_contribution}). The only significant difference between polydisperse and monodisperse occurs around the resonance in the dust density perturbation. Only when the resonance is completely out of the integration domain is growth possible again. This is because of the stabilizing effect of the contribution of the resonance, see (\ref{eq:si_res_cont}). In other words, growing modes exist when
\begin{align}
    \omega_0 - k_x u_x^{(0)}(\taus) = -\hat k_z\kappa - k_x \Delta u_x^{(0)}(\taus) < 0
\end{align}
for all $\taus$ in the size distribution. Since $\Delta u_x^{(0)} \approx -2\eta \taus$ for small $\stokes$ and $\mu$, this translates to $k_x > \hat k_z\kappa/(2\eta\taus)$. For the parameters of Figure \ref{fig:SI_contribution} ($\stokes_0=0.1$, $\Delta=0.3$ and $K_z=0.8$), we find that for $K_x>2.32$ the resonance is completely out of the integration domain, which agrees with the point where the dashed blue curve cuts off on the left.  

The monodisperse SI disappears for large $K_x$ because the stabilizing contribution of the perturbation in drift velocity overtakes the destabilizing contribution of the dust density perturbation. This is where the green and orange curves cross in Figure \ref{fig:SI_contribution}. The limiting $K_x$ can be quantified by first focusing on the monodisperse case. The full expression for $\omega_1$ from eigenvalue pertubation theory is not very enlightening, but we can gain some insight in the limit of small $\taus$, where we ignore terms that are quadratic in $\taus$:
\begin{align}
\frac{\omega_1}{\mu} =& \frac{k_x(2\Omega\rmi  \Delta u_y^{(0)} \pm \hat k_z \kappa  \Delta u_x^{(0)})}{2 \omega_{\Delta}}\left[1+ 3\rmi \taus  \omega_{\Delta}\right]\nonumber\\
-&  \omega_{\Delta} - \rmi \taus  \omega_{\Delta}^2 - \frac{\rmi \kappa^2}{2}(1+\hat k_z^2)\taus \pm \hat k_z\kappa(1+2\rmi \taus  \omega_{\Delta}), 
\end{align}
with $\omega_\Delta=\pm \hat k_z\kappa - k_x \Delta u_x^{(0)}$. The first term on the right hand side is due to the dust density perturbation, while the rest of the terms are due to the perturbation in relative velocity. 

Focus on the imaginary part of $\omega_1$, plugging in the expression for $\omega_{\Delta}$:
\begin{align}
\frac{\Im(\omega_1)}{\mu}=&\frac{k_x \Omega  \Delta u_y^{(0)}}{\pm \hat k_z\kappa - k_x \Delta u_x^{(0)}}
\pm\frac{3\taus k_x\hat k_z \kappa  \Delta u_x^{(0)}}{2}\nonumber\\
-& \taus k_x^2 \Delta {u_x^{(0)}}^2
- \frac{(1-\hat k_z^2)\kappa^2\taus}{2} 
\end{align}
For small $\stokes$, $\Delta u_y^{(0)}=\kappa\taus |\Delta u_x^{(0)}|/2$:
\begin{align}
\frac{2\Im(\omega_1)}{\kappa^2\taus\mu}=&
\frac{\Omega}{\kappa}\frac{k_x  |\Delta u_x^{(0)}|/\kappa}{k_x |\Delta u_x^{(0)}|/\kappa\pm \hat k_z }
\mp 3 \hat k_z   k_x|\Delta u_x^{(0)}|/\kappa\nonumber\\
-& 2 k_x^2 |\Delta {u_x^{(0)}}|^2/\kappa^2
- (1-\hat k_z^2).
\end{align}
We have zero growth (i.e. $\Im(\omega_1)=0$) when
\begin{align}
\pm\frac{\Omega}{\hat k_z^2 \kappa}
+ 2 q^3
\pm 5 q^2
+2q \mp 1=0
\label{eq:cubic}
\end{align}
with $q=\frac{k_x|\Delta u_x^{(0)}|}{\kappa\hat k_z}$. If we take the Keplerian case $\kappa=\Omega$, we find that in the limit $\hat k_z \rightarrow 1$, the relevant solution $q=2$, which makes the limiting $K_x=10$, which agrees with the location of the right edge of the region of growth towards large $K_z$ in figure \ref{fig:SI_monoresonance}. While this edge roughly follows the constant $q$ curve, which has the same shape as the resonance condition, at smaller $K_z$ deviations start to occur. At $K_z=0.8$, the case studied in figure \ref{fig:SI_contribution}, $q=2$ corresponds to $K_x=2.8$, while the edge of the region of positive growth rates is located at $K_x=3.5$, as correctly predicted by the solution of the cubic (\ref{eq:cubic}). This means that we are far enough away from the resonance to use regular perturbation theory to predict the edge of growth. 

In the polydisperse case, we have an integration over stopping time:
\begin{align}
\Im(\omega_1)= \frac{1}{{\rho_{\rm g}^{(0)}}}\frac{\kappa^4\hat k_z^2}{2k_x^2}\int\sigma^{(0)} q\left[
\frac{q}{q\pm 1}
\mp 3 \hat k_z^2 q 
- 2 \hat k_z^2 q^2
- 1+\hat k_z^2\right]\rmd q,
\end{align}
where we again have taken $\kappa=\Omega$ and used $\Delta u_x^{(0)}=-2\eta \taus$ to make the integration over $q$ rather than $\taus$. The condition that growth vanishes is now given by an average over the size distribution for all terms. For the particular size distribution (\ref{eq:sizedistSI}), the limiting wave number is very close to the monodisperse case (see figure \ref{fig:SI_contribution}). For size distributions that are dominated by larger (smaller) stopping times, the limiting wave number will move to the left (right). Hence, including larger stopping times in the size distribution makes it more difficult for the SI to survive in the polydisperse case.  

A reasonable approximation to the maximum growth rate in the polydisperse case with size distribution (\ref{eq:sizedistSI}) can be obtained by taking the growth rate of the corresponding monodisperse case, but at the wave number where the resonance is just outside the integration domain. In figure \ref{fig:SI_contribution}, this corresponds to $K_x = 2.4$. We find that
\begin{align}
\frac{\Im(\omega_1)}{\Omega}= \frac{\mu\stokes_0}{2}\left[\frac{1-\Delta}{\Delta} + \frac{\hat k_z^2}{1-\Delta}\left(4-\Delta - \frac{2}{1-\Delta}\right)\right].
\label{eq:SI_growth}
\end{align}
For $\Delta\rightarrow 0$, the first term in square brackets blows up because we are approaching the resonance, rendering the analysis invalid. For $\Delta \rightarrow 1$, the last term dominates, stabilising the flow. For intermediate values of $\Delta$, the term in square brackets is $O(1)$, and we find a growth rate $\Im(\omega_1)/\Omega \sim \mu \stokes_0$, which is a factor $\sqrt{\mu}$ lower than the monodisperse case \citep{2018MNRAS.477.5011S}. Growth is no longer due to the resonance, but due to a regular perturbation of an inertial wave. From the approximate relation (\ref{eq:SI_growth}), we find that this intermediate regime is bounded by $\Delta < 0.5$. 

\begin{figure}
  \resizebox{\hsize}{!}{\includegraphics{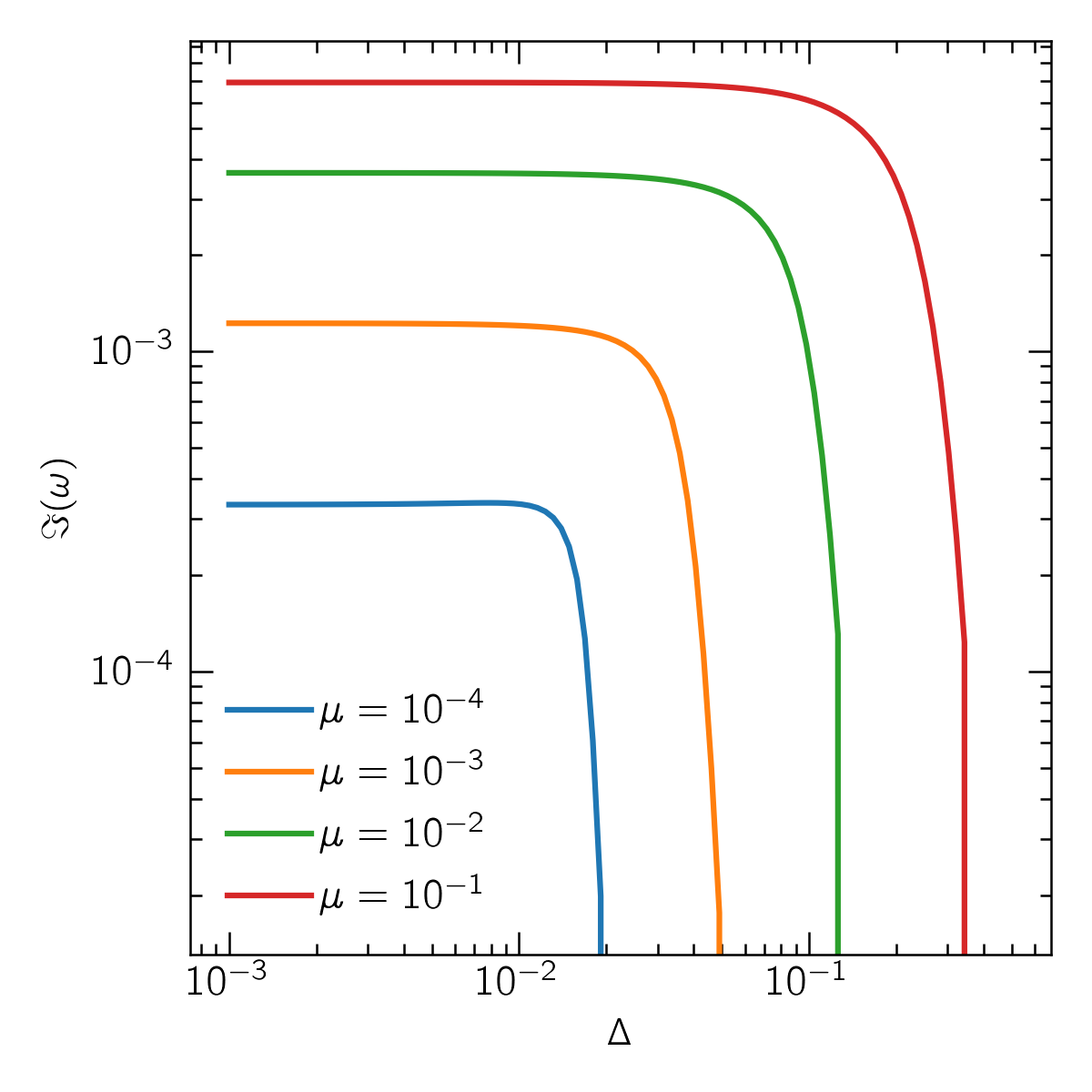}}
  \caption{Growth rates of the SI with $\stokes_0=0.1$ and varying width of the size distribution, at $K_x=2$ and $K_z=0.85$, which in the monodisperse case is the location of the resonance.}
  \label{fig:SI_delta}
\end{figure}

As with the acoustic drag instability (see Paper I), for size distributions that are narrow enough there exists the quasi-monodisperse regime, where the system acts as if the dust is monodisperse. How narrow the size distribution needs to be for this to happen is illustrated in figure \ref{fig:SI_delta}. The wave numbers are chosen such that we are at the resonance in the monodisperse case, and towards smaller values of $\Delta$ it is clear that growth rates are $\propto \sqrt{\mu}$. For wider size distributions growth vanishes at the location of the resonance, until at $\Delta=0.4$ we do not see any growth even for $\mu=0.1$. Since the real part of the monodisperse frequency perturbation is given by $\hat k_z \sqrt{\mu/2}\Omega$ \citep{2018MNRAS.477.5011S}, we find that the quasi-monodisperse regime exists for 
\begin{align}
    \Delta < \frac{\hat k_z \sqrt{\mu}}{2\sqrt{2} K_x \stokes_0}.
\end{align}
If we restrict ourselves to the resonance corresponding to $\stokes_0$, this becomes
\begin{align}
    \Delta < \sqrt{\frac{\mu}{2}}.
    \label{eq:delta_SI}
\end{align}
In figure \ref{fig:SI_delta}, this corresponds to where the curves are horizontal. So while the quasi-monodisperse regime is more accessible for smaller stopping times and smaller horizontal wave numbers, the condition for resonant growth only depends on the dust to gas ratio, see (\ref{eq:delta_SI}). This agrees with \cite{2020MNRAS.499.4223P}, whose case with $\mu=0.5$ and a wide size distribution showed no growth at resonant wave numbers (their figure 5).  

\subsection{Summary}

The resonance that gives rise to the SI in the monodisperse regime is an enemy of polydisperse growth. Only for very narrow size distributions do we recover the monodisperse regime, see (\ref{eq:delta_SI}). For slightly wider size distributions, integrating over the resonance regularizes the perturbation, making the SI no longer an RDI. Despite the stabilizing effect of the resonance, a band of unstable wave numbers can persist for size distributions of moderate width, with growth rates $\sim \mu \stokes_0$ that are non-resonant. From (\ref{eq:SI_growth}), for this band to persist for all $\hat k_z$ requires $\Delta < 0.5$. For wide size distributions including particles of very small Stokes numbers the SI switches off completely for $\mu \ll 1$, an effect which is captured in (\ref{eq:SI_growth}). Therefore, the polydisperse, low-$\mu$ SI is unlikely to play a role in realistic settings of planet formation. The non-resonant, high-$\mu$ variant of the SI, which we do not study here, is more promising in the polydisperse regime, if the size distribution is favourable \citep{2021MNRAS.502.1469M}.

\section{Settling Instability}
\label{sec:settling}

We now let $z_0 \neq 0$ and study the settling instability. While the streaming instability has received more attention in the context of planet formation, the settling instability is interesting for several reasons: it generally grows faster than the SI, can grow fast in the RDI regime of $\mu \ll 1$, and growth rates can be independent of stopping time \citep{2018MNRAS.477.5011S}. Moreover, it does not seem to be affected by having a distribution of dust sizes, and convergence with number of dust species is much easier achieved than for the SI  \citep{2020MNRAS.497.2715K}. 

\begin{figure}
  \resizebox{\hsize}{!}{\includegraphics{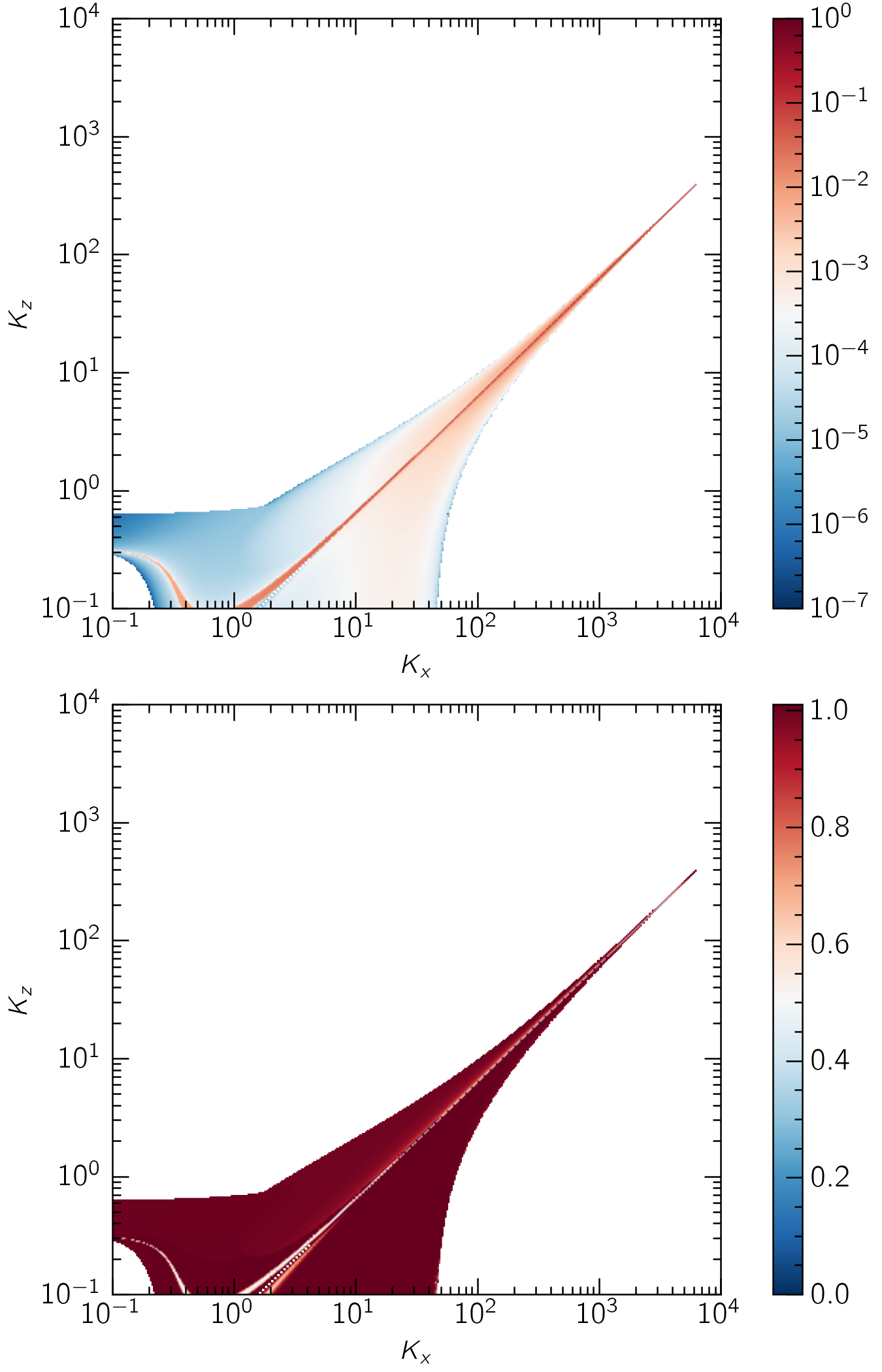}}
  \caption{Top panel: growth rates for the monodisperse settling instability with $\stokes=0.1$, $\mu=0.0001$ and $z_0/H=1$. Bottom panel: $\rmd\log \Im(\omega)/\rmd\log\mu$ at $\mu=0.0001$.}
  \label{fig:dsi_monoresonance}
\end{figure}

We start off by setting $z_0=H = c/\Omega$, where $H$ is a measure of the vertical thickness of the gas disc. This is also the value considered in \cite{2020MNRAS.497.2715K}. Note that this implies a settling velocity that is larger than the radial drift velocity by more than an order of magnitude:
\begin{align}
    \frac{\Delta u_z}{\Delta u_x} \approx \frac{z_0\Omega^2\taus}{2\eta\taus} = \frac{z_0}{H} \frac{c\Omega}{2\eta} \approx 16 \frac{z_0}{H},
\end{align}
where in the last step we have used our choice of $\eta/(c\Omega)=10^{-3/2}$. We also work with a smaller than usual dust to gas ratio, $\mu=10^{-4}$. This is because due to the larger growth rates of the settling instability, higher values of $\mu$ yield perturbations in the eigenvalues that are no longer small (basically because $\tilde z_0 \gg 1$). Choosing such a small value of $\mu$ ensures that we remain safely in the regime where resonant drag theory is fully valid. 

In figure \ref{fig:dsi_monoresonance}, we show growth rates for the monodisperse DSI with $\stokes=0.1$. The resonance clearly stands out as the red curve with highest growth rates (a similar pattern was shown in \cite{2020MNRAS.497.2715K} for $\stokes=0.01$), which even for $\mu=10^{-4}$ easily reach $\Im(\omega)/\Omega=0.01$. In the bottom panel of figure \ref{fig:dsi_monoresonance}, the resonance stands out as the white curve, showing slower growth with $\mu$ compared to the red regions of regular (non-resonant) perturbations. 

\begin{figure}
  \resizebox{\hsize}{!}{\includegraphics{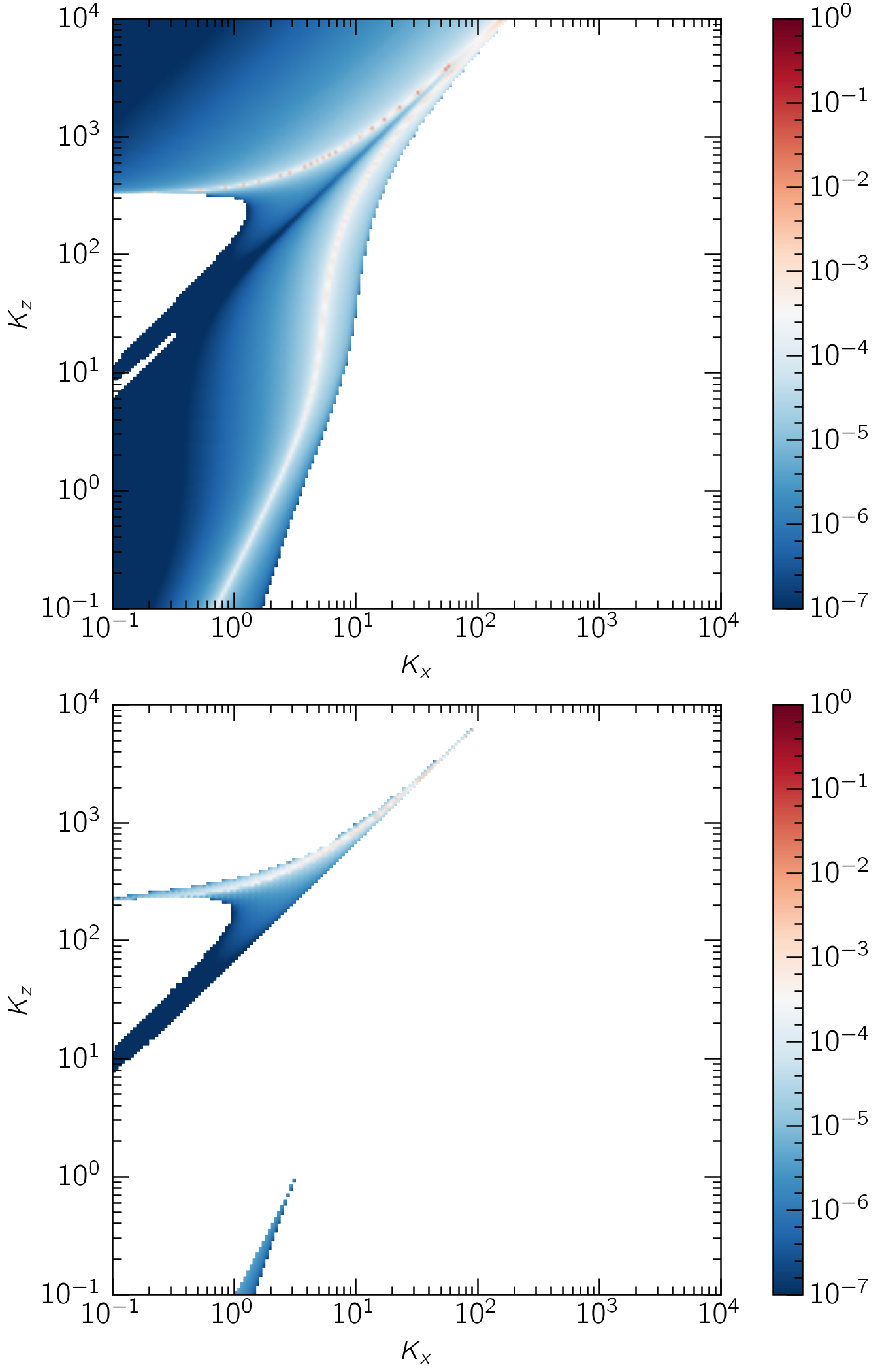}}
  \caption{Growth rates for the settling instability for $\mu=10^{-4}$ and $z_0/H=0.001$. Top panel: monodisperse with $\stokes=0.1$. Bottom panel: polydisperse with $\stokes_0=0.1$ and size distribution width $\Delta=0.5$.}
  \label{fig:dsi_contour_growth}
\end{figure}

Towards the bottom of figure \ref{fig:dsi_monoresonance}, the white curve splits into two branches. This becomes more obvious in this wave number range if we consider smaller values of $z_0$. In figure \ref{fig:dsi_contour_growth}, we show growth rates for $\stokes_0=0.1$ and $\mu=10^{-4}$, similar to figure \ref{fig:dsi_monoresonance}, but at $z_0/H=0.001$. The top panel shows results for the monodisperse case, and the resonant branches show up as white regions of high growth. The bottom branch, hardly visible in figure \ref{fig:dsi_monoresonance}, and in fact not shown in figure 1 of \cite{2020MNRAS.497.2715K}, is the branch that in the limit of $z_0 \rightarrow 0$ gives rise to the streaming instability. Towards small $K_z$, it already maps onto the SI resonant location seen in figure \ref{fig:SI_monoresonance}, but note that the growth rates are smaller due to the smaller value of $\mu$ considered in figure \ref{fig:dsi_contour_growth}. 

In the bottom panel of figure \ref{fig:dsi_contour_growth}, the polydisperse case with a constant size distribution with $\stokes/\stokes_0 \in [0.5,1.5]$ is considered. Here we see that the SI branch suffers its usual fate of quenching under a size distribution. As in the pure SI case, integrating over the resonance destroys growth. On the other hand, the resonance due to settling survives a size distribution. In this case, since it is the other inertial wave that creates the resonance, the resonance has a positive contribution to growth in the polydisperse case. This can be seen from (\ref{eq:res_cont}), as the vertical settling is in resonance with the inertial wave that has the opposite frequency compared to the radial drift. As a result, this branch survives relatively unscathed even for relatively wide size distributions.

\begin{figure}
  \resizebox{\hsize}{!}{\includegraphics{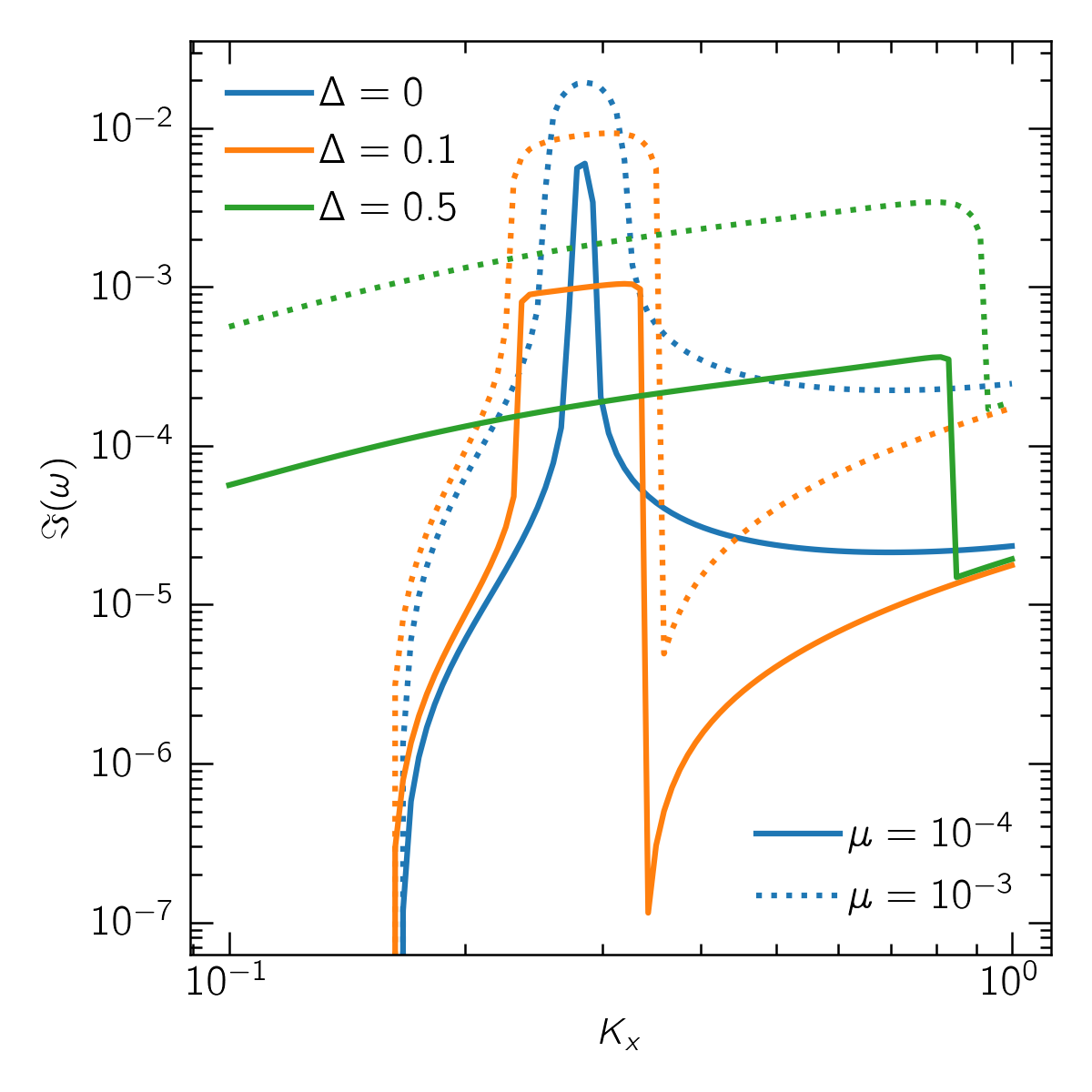}}
  \caption{Growth rates for the DSI with $z_0/H=1$ and $\stokes_0=0.1$, at fixed vertical wavenumber $K_z=0.2$. Solid curves: $\mu=10^{-4}$, dashed curves: $\mu=10^{-3}$. Different colors indicate different widths of the size distribution. }
  \label{fig:dsi_poly_kx}
\end{figure}

This is further illustrated in figure \ref{fig:dsi_poly_kx}, where we return to $z_0/H=1$ and fix $K_z=0.2$. From figure \ref{fig:dsi_monoresonance}, we see that for this value of $K_z$, we cross two resonances at $K_x\approx 0.3$ and $K_x\approx 3$. In the latter case, the two branches of resonance come together into what was called the double resonant angle in \cite{2018MNRAS.477.5011S}, which is the subject of the next subsection. The resonance at $K_x\approx 0.3$ is only active for $z_0\neq 0$, and in figure \ref{fig:dsi_poly_kx} the blue curves, indicating the monodisperse case, clearly show the growth rate peak at the resonance. Moreover, when changing the dust to gas ratio by a factor of 10 (solid vs. dotted curves), growth at the resonance changes only by a factor of $\sqrt{10}$, the hallmark of a resonant drag instability. 

The polydisperse cases shown in figure \ref{fig:dsi_poly_kx} have a constant size distribution with $\stokes/\stokes_0 \in [1-\Delta, 1+\Delta]$, as usual. For $\Delta=0.1$, the resonance can still be identified, but for $\Delta=0.5$ it is smeared out considerably. Importantly, growth rates are now everywhere $\propto \mu$: even for a relatively narrow size distribution with $\Delta=0.1$, there is a factor of 10 difference in growth rate between the solid and dotted orange curves. This shows that the perturbation is now regular, i.e. no longer resonant. Furthermore, we expect from (\ref{eq:res_cont}) that the contribution of the resonance $\propto 1/\Delta$, which agrees with the differences between $\Delta=0.1$ and $\Delta=0.5$ around $K_x=0.3$.

The difference in maximum growth rate around the resonance between the monodisperse case and the polydisperse case will therefore strongly depend on $\mu$, since the former scales as $\sqrt{\mu}$ and the latter as $\mu$. In figure \ref{fig:dsi_poly_kx}, for $\mu=10^{-4}$ the maximum monodisperse growth rate is a factor of 20 larger than the polydisperse case with $\Delta=0.5$, but for $\mu=10^{-3}$ is is only a factor of $5.5$.  

Further experiments with narrower size distributions showed that for $\Delta < 0.01$, we enter the quasi-monodisperse regime for $\mu = 10^{-4}$. This is consistent with the requirement $\Delta \lesssim \sqrt{\mu}$, similar to the SI (see (\ref{eq:delta_SI}). 

\subsection{Double resonance}

One of the interesting features of the DSI is that growth rates approach infinity at the double resonant angle where ${\bf k}\cdot \Delta{\bf u} = 0$. This is where the two resonant branches merge, as seen in the top panel of figure \ref{fig:dsi_contour_growth}.

If we want the resonance to behave in a quasi-monodisperse fashion, we again need that the spread in ${\bf k}\cdot \Delta{\bf u}$ due to the size distribution is much smaller than the perturbed gas wave velocity, which for the double resonance is $\omega_{\rm mono}/\Omega=-(2\stokes\mu K_x)^{1/3}/2$ \citep{2018MNRAS.477.5011S}. In the limit of $\mu\rightarrow 0$, the relative velocities are
\begin{align}
    \Delta u_x =& -\frac{2\eta}{\Omega} \frac{\stokes}{1+\stokes^2},\\
    \Delta u_z =& \Omega z_0 \stokes.
\end{align}
If we take a narrow size distribution with $\stokes/\stokes_0 \in [1-\delta/2, 1+\delta/2]$, we find that to first order in $\delta$, we get a spread of
\begin{align}
  \frac{\Delta({\bf k}\cdot \Delta{\bf u})}{\Omega} = \left[K_z \tilde z_0 \stokes_0-2K_x\frac{\stokes_0}{1+\stokes_0^2}\right]\delta +\frac{4K_x\stokes_0^3\delta}{(1+\stokes_0^2)^2}.
  \label{eq:dku}
\end{align}
The term in square brackets is zero at the resonant Stokes number $\stokes_0$. For a narrow size distribution, $\delta \ll 1$, one can develop a series in $\delta$, with the monodisperse result given by $\omega_{\rm mono}$. In order to stay close to the monodisperse regime, i.e. quasi-monodisperse, we require that integrals of the form 
\begin{align}
\int \frac{\sigma^{(0)}f(\stokes)}{\omega_{\rm mono} - {\bf k}\cdot \Delta {\bf u}^{(0)}}\rmd \stokes,
\end{align}
form a converging series in $\delta$. With help of (\ref{eq:dku}) and the expression for $\omega_{\rm mono}$ evaluated close to $\stokes_0$, we readily find that we need
\begin{align}
    \delta \ll \left(\frac{1}{3} + \frac{8K_x\stokes_0^3}{\left(1+\stokes_0^2\right)(2\stokes_0 \mu K_x)^{1/3}}\right)^{-1}.
    \label{eq:dsi_lim_delta}
\end{align}

\begin{figure}
  \resizebox{\hsize}{!}{\includegraphics{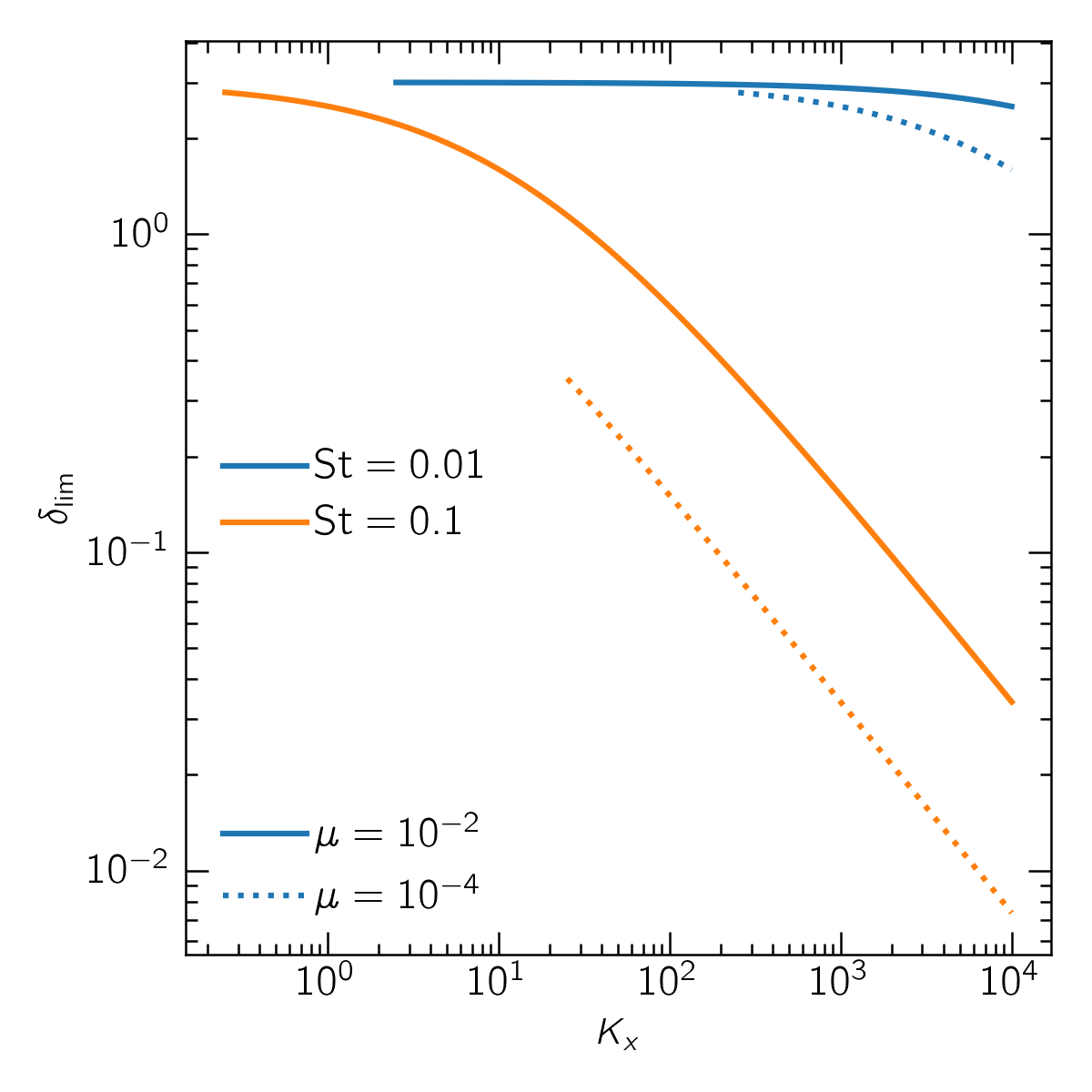}}
  \caption{Limiting width of a constant size distribution with $\stokes/\stokes_0 \in [1-\delta, 1+\delta]$, see (\ref{eq:dsi_lim_delta}), for different central Stokes numbers and dust to gas ratios.}
  \label{fig:dsi_lim_delta}
\end{figure}

The quasi-monodisperse is most easily reached for small $K_x$, small Stokes number, and high dust-to-gas ratio. Some examples are shown in figure \ref{fig:dsi_lim_delta}. When $\delta_{\rm lim} \sim 1$, growth rates at the double resonance will be unaffected by even a wide size distribution. As discussed in \cite{2018MNRAS.477.5011S}, the double resonance only plays a role for 
\begin{align}
K > \frac{\left(\frac{\pi}{2}-|\theta_{\rm res}|\right)^3}{\mu\stokes},
\end{align}
where $\theta_{\rm res} = \arctan(\Delta u_z/\Delta u_x)$ is the double resonant angle. This is the wave number where the curves in figure \ref{fig:dsi_lim_delta} are cut-off on the left. From the figure, we see that the smaller Stokes number setup never leaves the quasi-monodisperse regime: we expect to see similar growth rates for polydisperse dust to the monodisperse case, even for relatively wide size distributions. Due to the strong Stokes number dependence of (\ref{eq:dsi_lim_delta}), the situation is very different for $\stokes_0=0.1$. For $\mu=10^{-4}$, the quasi-monodisperse regime is out of reach for wide size distributions for all wave numbers, while for $\mu=10^{-2}$ it is attainable only for small $K_x$. Importantly, the DSI shows highest growth rates for large $K_x$ \citep{2018MNRAS.477.5011S}. 

\begin{figure}
  \resizebox{\hsize}{!}{\includegraphics{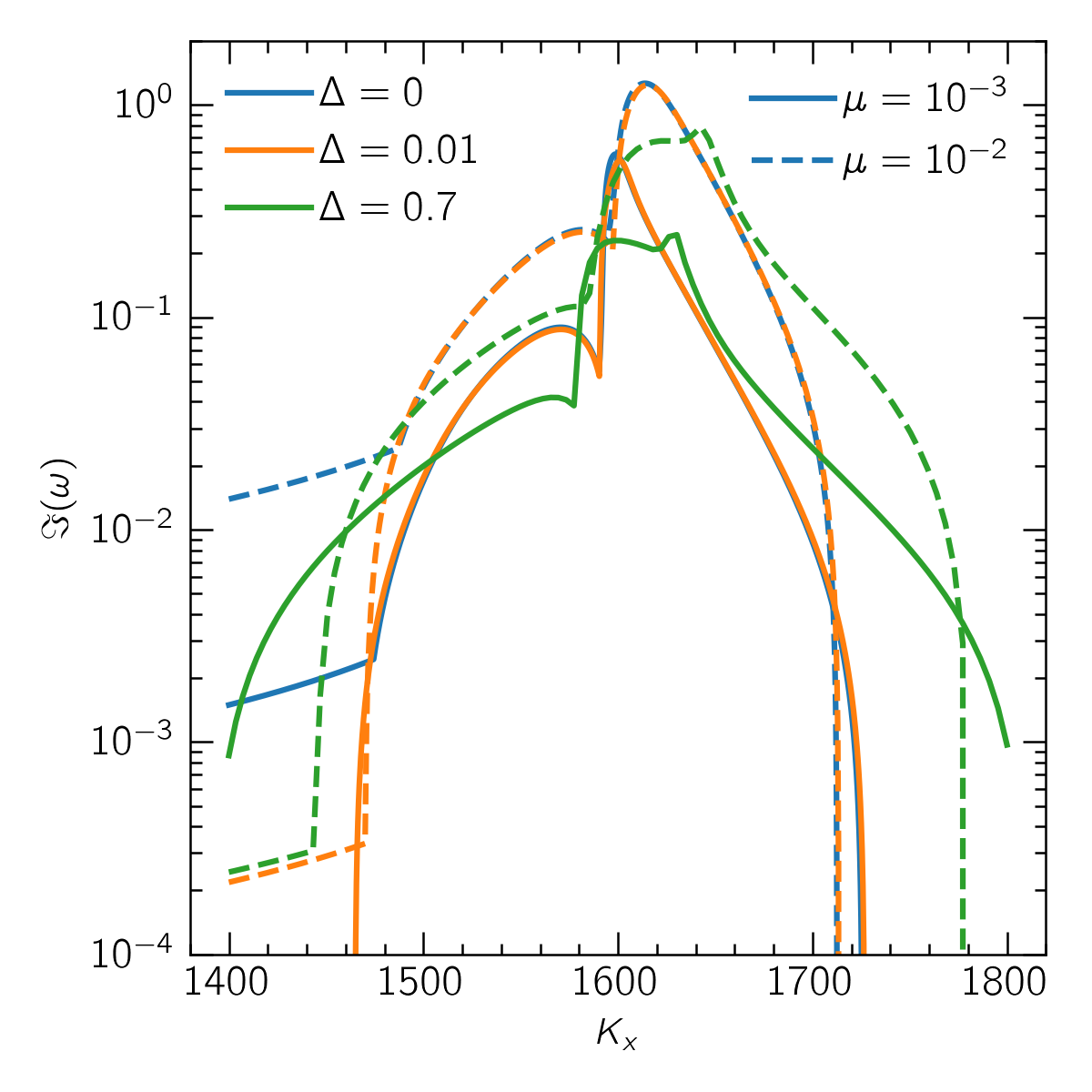}}
  \caption{Growth rates for the DSI at $z_0=H$, $\stokes_0=0.1$ and $K_z=100$. Different colors indicate different widths of the size distribution, and we consider $\mu=10^{-3}$ (solid curves) and $\mu=10^{-2}$ (dashed curves). The location of the double resonance in the monodisperse regime, in the limit $\mu=0$ and $\stokes \rightarrow 0$, is at $K_x\approx 1580$.}
  \label{fig:dsi_double_growth}
\end{figure}

We show numerically calculated growth rates around the double resonance in figure \ref{fig:dsi_double_growth}. The blue curves, indicating monodisperse results, being a factor 10 apart in $\mu$, show a factor of 10 difference in growth rate at the left end of the plot, indicating non-resonant growth. At the growth peaks, they differ only by a factor $2.17 \approx 10^{1/3}$, clearly indicating growth $\propto \mu^{1/3}$ at the double resonant angle.

If we take a size distribution with Stokes numbers between $\stokes_0(1-\Delta)$ and $\stokes_0(1+\Delta)$, with $\Delta < 1$ but not necessarily small, we expect from figure \ref{fig:dsi_lim_delta} that the quasi-monodisperse regime will be confined to $\Delta \lesssim 0.1$ at $K_x\approx 1500$. Indeed, the narrow size distribution with $\Delta =0.01$ (orange curve in figure \ref{fig:dsi_double_growth}) closely follows the monodisperse case around the resonance (but not towards the non-resonant regime at the left edge of the figure). A wider size distribution with $\Delta=0.7$ (green curves in figure \ref{fig:dsi_double_growth} still shows the highest growth rates close to the resonant location, but growth is reduced compared to the monodisperse case, and no longer $\propto \mu^{1/3}$. In fact, it appears that the maximum growth rate $\propto \mu^{1/2}$, but more work is necessary to establish the exact dependence on $\mu$.

It should be noted that the large wave numbers that are the domain of the double resonance can be problematic for the fluid approximation for the dust, and moreover that they are particularly susceptible to viscous damping, which we explore below.

\subsection{Gas viscosity and dust diffusion}

\begin{figure*}
  \resizebox{\hsize}{!}{\includegraphics{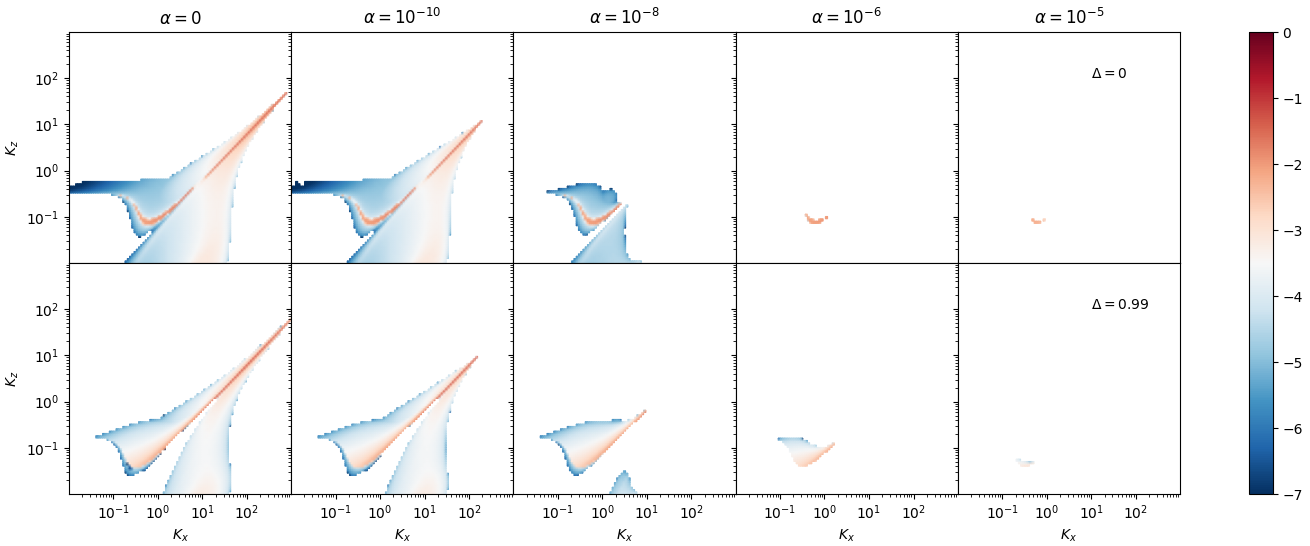}}
  \caption{Growth rates for $\mu=10^{-4}$ and $z_0=H$ for different levels of viscosity. The top row shows monodisperse results; the bottom row polydisperse with a wide size distribution ($\Delta=0.99$).}
  \label{fig:dsi_visc}
\end{figure*}

\begin{figure*}
  \resizebox{\hsize}{!}{\includegraphics{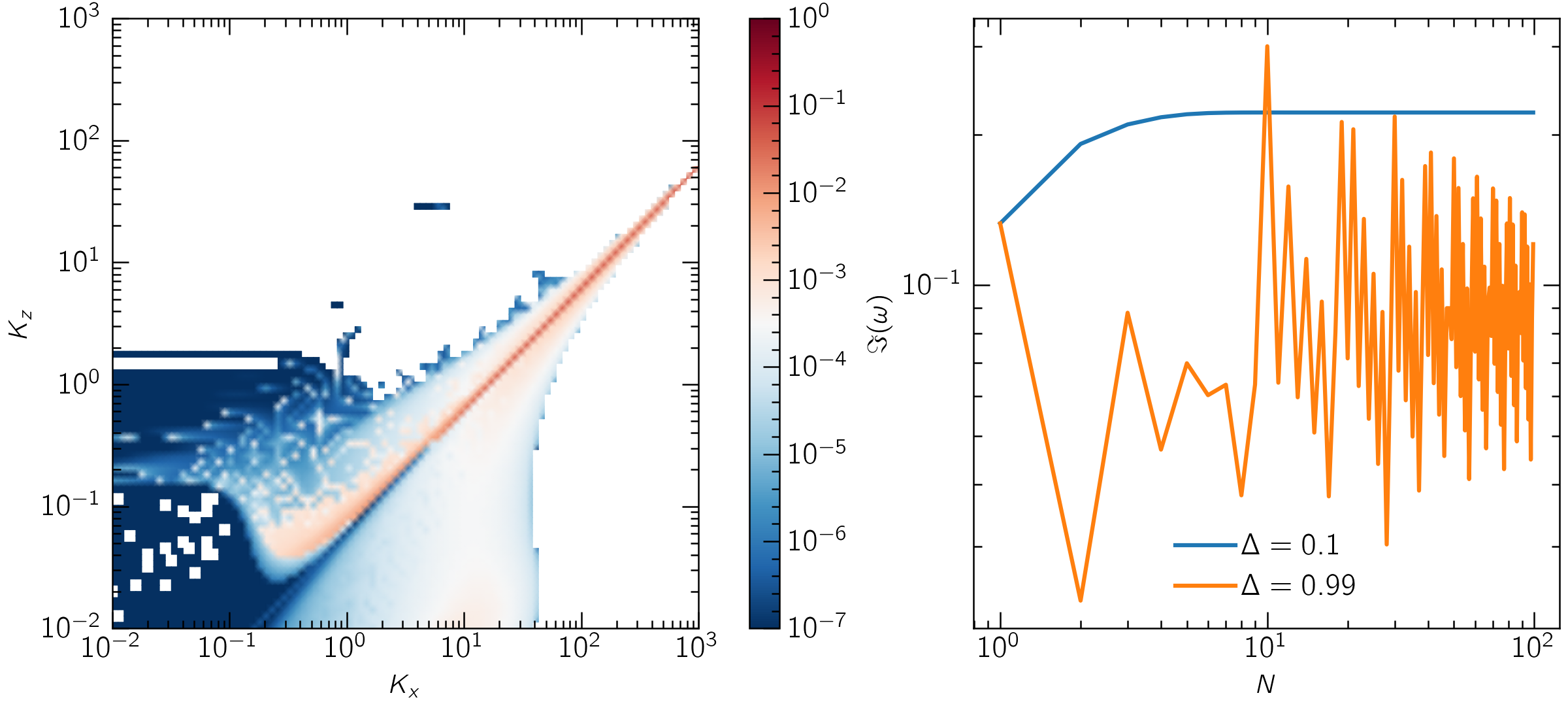}}
  \caption{Left: growth rates for $\mu=10^{-4}$ and $z_0=H$, calculated with $N=100$ discrete dust sizes sampling a size distribution with $\Delta=0.99$. Right: convergence with $N$ close to the double resonant angle with $K_x=16007$ and $K_z=1000$ for $\Delta=0.1$ and $\Delta=0.99$.}
  \label{fig:dsi_convergence}
\end{figure*}

We finally add gas viscosity and dust diffusion to the problem, in the same way as in \cite{2021MNRAS.502.1469M}, following the formulation of \cite{2020ApJ...891..132C}. As mentioned in \cite{2021MNRAS.502.1469M}, this model does not capture turbulent clumping, or the effect of dust on gas turbulence, but only represents average effects of the turbulence. With this in mind, we show growth rates for the DSI for $z_0=H$ and $\mu=10^{-4}$ for varying levels of viscosity in figure \ref{fig:dsi_visc}. 

From the inviscid case (leftmost panels in figure \ref{fig:dsi_visc}), we see the disappearance of the SI branch of the resonance, and the smearing out of the DSI-related resonance. Towards larger $K_x$, we see that the double resonance is far less affected, indicating that the maximum growth rate in the polydisperse regime will be comparable to the monodisperse case. 

If we start to increase gas viscosity and associated dust diffusion, we see that the double resonance disappears first (middle panels of figure \ref{fig:dsi_visc}), due to the fact that it is confined to large wave numbers and therefore more susceptible to diffusion. This is also where the maximum growth rate starts to differ between the monodisperse and polydisperse case. For $\alpha=10^{-5}$, the difference is almost an order of magnitude.   

\subsection{Numerical convergence in the DSI}
\label{sec:appendix_conv}

A different approach than presented in \cite{2021MNRAS.502.1579P} for solving roots of the dispersion relation (\ref{eq:dsi_dispersion}), which involves integrals over stopping time, is to first discretize the backreaction on the gas:
\begin{align}
    \frac{\rmi}{{\rho_{\rm g}^{(0)}}}\int \frac{\sigma^{(0)}}{\taus}\left[\Delta {\bf u}^{(0)}\frac{\hat\sigma}{\sigma^{(0)}} + {\bf \hat u} - {\bf \hat v}\right]\rmd \taus\equiv &\int f(\taus)\rmd \taus \nonumber\\
    \approx &\sum_n f(\tau_{{\rm s},n})w_n,
\end{align}
with nodes $\tau_{{\rm s},n}$ and weights $w_n$. For wide size distributions, it is convenient to perform the integral in log space. There is further freedom in choosing equidistant nodes \citep{2019ApJ...878L..30K, 2020MNRAS.497.2715K}, or Gauss-Legendre nodes. For $N$ integration nodes, we end up with a straightforward eigenvalue problem with matrix size $(4 + 4N)\times (4 + 4N)$, that can be solved by standard methods. This was termed the 'direct solver' in \cite{2021MNRAS.502.1469M}.

It was shown in \cite{2019ApJ...878L..30K} and \cite{2021MNRAS.502.1579P} that for the streaming instability, in large regions of parameter space this method fails to converge, or converges very slowly with $N$. For $N \gtrsim 4000$, the problem quickly becomes computationally intractable. The problem can be traced to the nastiness of the integrand \citep{2021MNRAS.502.1579P} that is close to singular. Not only does that mean that linear growth rates are more difficult to obtain, there is a danger that numerical simulations (that can not afford to have 10000s of dust species) will pick up spurious growing modes that come to dominate the simulation. 

In \cite{2020MNRAS.497.2715K}, it was shown that the situation looks better for the settling instability: the maximum growth rate over a range of wave numbers $(k_x,k_z) \in 2\pi [1,1000]/H$ converges for $N \gtrsim 32$ with $\mu=0.01$ and $z_0/H=1$. Here, we present some additional numerical experiments, with some more details on where and when we can expect fast convergence. 

\begin{figure}
  \resizebox{\hsize}{!}{\includegraphics{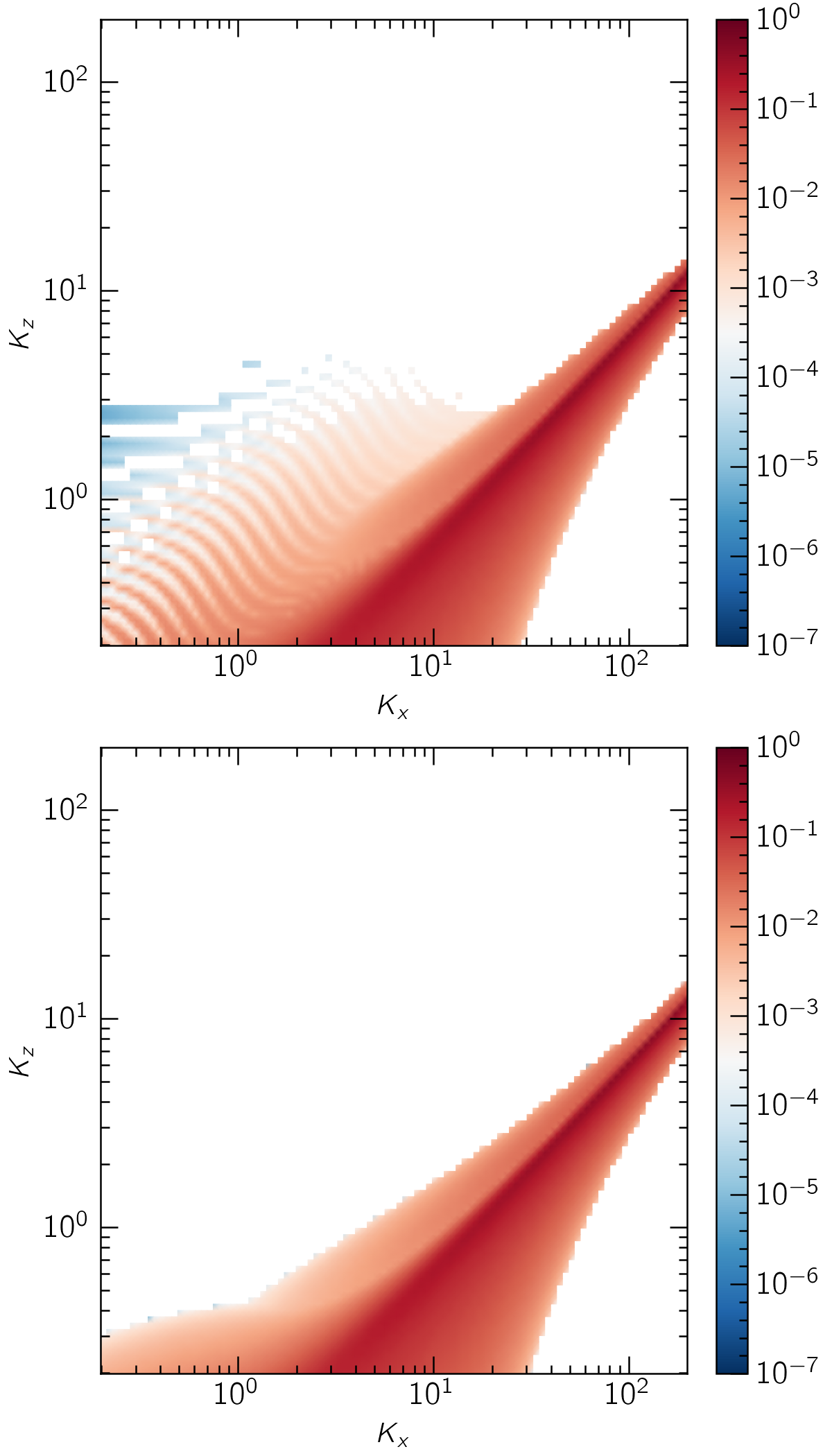}}
  \caption{Growth rates for the DSI with $\mu=0.01$, $\stokes_0=0.1$ and a constant size distribution with $\Delta=0.99$. Top panel: calculation using the direct method, using $32$ dust species, equidistant in log space. Bottom panel: calculation using the method of \cite{2021MNRAS.502.1579P}.}
  \label{fig:dsi_convergence_max}
\end{figure}

In the left panel of figure \ref{fig:dsi_convergence} we show growth rates calculated using the direct solver with $N=100$. This panel can be compared to the bottom left panel of figure \ref{fig:dsi_visc}. It is obvious that there is a lot of numerical noise, in particular towards the lower left part of the figure. The dotted structure around $K_x=K_z=0.5$ are artifacts resulting from integrating over the DSI-related resonance. Note also that the streaming instability branch is present, even though it should not be (see figure \ref{fig:dsi_visc}). This is again due to integration errors over the resonance \citep{2019ApJ...878L..30K, 2021MNRAS.502.1579P}. 

In the right panel of figure \ref{fig:dsi_convergence}, we show a convergence study at fixed wave numbers around the double resonant angle. For a relatively narrow size distribution, convergence is obtained even for $N=10$, while for the wider size distribution, numerical errors prevent convergence. It seems as though the amplitude of the noise goes down with $N$, so perhaps convergence can be reached at much larger $N$ than considered here.

The situation improves if we increase the dust to gas ratio to $\mu=0.01$. In particular, this takes the double resonant angle into the quasi-monodisperse regime, which is easier for standard integration methods to deal with. A case similar to figure \ref{fig:dsi_convergence_max} was studied in \cite{2020MNRAS.497.2715K} (their figure 2), who showed that the maximum growth rate converges for $N \geq 32$. They used a different size distribution, but we also find that the maximum growth rate, which occurs towards the double resonant angle at the highest $K_x$, does not change significantly upon increasing $N$. However, the top panel of figure \ref{fig:dsi_convergence_max} shows numerical artifacts for smaller $K_x$, similar to those seen in figure \ref{fig:dsi_convergence}. Again, these are caused by integrating over the resonance using a method that is not suited to deal with near-singularities. This is something to be aware of in numerical simulations of the DSI, that, as is the case for the SI, spurious modes may grow, albeit slower than the double resonant modes if these are present. 

\subsection{Summary}

The polydisperse settling instability differs from the polydisperse streaming instability in two important ways. First of all, the resonant branch associated with the settling velocity gives a \emph{positive} contribution when integrated over. This means that, unlike the SI, the polydisperse DSI survives in the limit of $\mu \ll 1$. Integrating over the resonance regularizes the perturbation, so that the growth rate is now $\propto\mu$ rather than $\propto \sqrt{\mu}$ as in the monodisperse case. 

Second, at the double resonant angle, growth can be less sensitive to having a size distribution, at least for maximum Stokes numbers that are much smaller than unity (see figure \ref{fig:dsi_lim_delta}). In those cases, the perturbation to the gas wave is strong enough that the instability remains in the quasi-monodisperse regime. Unfortunately, as the double resonance is confined to relatively large wave numbers, it is more susceptible to diffusion.

\section{Discussion and conclusion}
\label{sec:conclusion}

We have presented linear calculations of two resonant drag instabilities in the polydisperse regime that are thought to be important in protoplanetary discs: the streaming instability and the settling instability. The results can provide a starting point for nonlinear simulations, which will in the end decide if these instabilities for example lead to clumping and planetesimal formation \citep[e.g.][]{2020MNRAS.497.2715K}. We will consider this in a forthcoming paper. 

Several important simplifications were made to make the problem tractable. First of all, we have only considered the limit $\mu \ll 1$. It is important to note that this is not the regime usually studied for the SI, where it is usually assumed that for example dust settling has led to a dust-to-gas ratio of order unity. This regime was studied in the polydisperse case for example in \cite{2019ApJ...878L..30K, 2021MNRAS.501..467Z, 2021MNRAS.502.1469M}. For $\mu \ll 1$, the SI is a resonant drag instability, while it changes character for $\mu>1$ \citep{2018MNRAS.477.5011S}. By keeping $\mu\ll 1$, we can make firm contact with RDI theory, leading to a deeper understanding of the polydisperse SI.

We have only considered very simple (i.e. constant) size distributions. The dependence on form of the size distribution on the growth rates is different for each instability. Since the contribution of the resonance to the SI is negative, and the dust density contribution to the growth rate is dominated by the resonance, it is virtually impossible to pick a size distribution that leads to growth. Basically, one would have to exclude the 'resonant size' completely from the size distribution. For larger values of $\mu$, the resonance is less dominant, and there the form of the size distribution can have an impact \citep{2021MNRAS.502.1469M}. A similar story holds for the DSI, but in a positive way: since the contribution of the resonance is positive, it will be hard to find a size distribution that reduces the growth rates significantly compared to the constant size distribution considered here. For reference, for the acoustic drag instability, the contribution of the resonance can be positive or negative, depending on the asymmetry of the size distribution with respect to the resonant size (see Paper I).

It may seem paradoxical that to get growth for the PSI, the corresponding 'resonant size' needs to be excluded from the size distribution, even though the resonance is responsible for growth in the SI. The following thought experiment might help clarify why this is the case. Consider the monodisperse streaming instability at wave number ${\bf k}_{\rm mono}$ and Stokes number ${\rm St}_{\rm mono}$, and eigenvalue $\omega_{\rm mono}\approx \omega_{\rm inertial}=k_{x,{\rm mono}}v_{{\rm g}x}^{(0)}-\hat k_{z,{\rm mono}}\kappa$. Take the wave number to meet the resonant condition so that $\omega_{\rm inertial}={\bf k}_{\rm mono}\cdot {\bf u}^{(0)}$, and the dust to gas ratio to be $\mu_{\rm mono}\ll 1$. Now add a polydisperse component with $\mu_{\rm poly}\ll\mu_{\rm mono}$. Because $\mu_{\rm poly}\ll\mu_{\rm mono}$, the polydisperse component is not able to appreciably change the eigenvalue from $\omega_{\rm mono}$. Now consider the size density perturbation in the polydisperse component:
\begin{align}
    \frac{\hat\sigma}{\sigma^{(0)}} = \frac{{\bf k}_{\rm mono}\cdot {\bf\hat u}}{\omega_{\rm mono}-{\bf k}_{\rm mono}\cdot {\bf u}^{(0)}}.
    \label{eq:size_density_pert}
\end{align}
This term, multiplied by $\Delta {\bf u}^{(0)}$, enters the backreaction ${\bf \hat b}$ under the integral sign, and is the main driver of instability. Assume for the sake of argument that ${\bf k}_{\rm mono}\cdot {\bf\hat u}>0$ for all $\stokes$. The denominator is 
\begin{align}
\omega_{\rm mono}-k_{x,{\rm mono}} u_x^{(0)}\approx 
-\hat k_{z,{\rm mono}}\kappa + 2\eta\tau_{\rm s}k_{x,{\rm mono}},
\end{align}
which is an odd function of $\taus$ around the resonant stopping time. This means that Stokes numbers smaller than $\stokes_{\rm mono}$ will have a \emph{positive} contribution to the backreaction (note that the multiplication with $\Delta {\bf u}^{(0)}$ changes the sign), while Stokes numbers greater than $\stokes_{\rm mono}$ will have a \emph{negative} contribution to the backreaction. The closer to the resonance, the stronger these contributions are. This means that, on one side of the resonance, Stokes numbers arbitrarily close to the resonance have a strongly negative effect on the growth rate of the instability, even though growth is caused by the monodisperse component at the resonance.

The reason for this apparent paradox is that the polydisperse component is forced at the wrong frequency everywhere except exactly at the resonance. The total effect of the polydisperse component can be found by integration over the size distribution. Whether the resonance has a positive or negative contribution now crucially depends on the numerator in the size density perturbation (\ref{eq:size_density_pert}) and how it depends on Stokes number. This causes the difference between the SI resonance and the DSI resonance, where in the case of the SI the resonance damps growth, and in the case of the DSI it promotes growth. In reality, we do not have dust made up of a dominant monodisperse component and a polydisperse component of negligible mass, but the issue remains that most of the polydisperse dust will be forced at the wrong frequency.

A very important simplification is that we have considered only unstratified shearing boxes. For the SI, this restricts the analysis to close to the mid plane, while for the DSI it means that we can only consider time scales that are short compared to the settling time. The fully stratified case is more complex to deal with, because setting up an equilibrium requires including a turbulence model to keep the dust from settling \citep{2021ApJ...907...64L}. An unstratified model can be used to delineate effects of diffusion, but it is clear that future models should include stratification, to bring linear results in closer contact with numerous nonlinear simulations \citep[see e.g.][for recent results]{2021ApJ...919..107L, 2021ApJ...911....9K, 2024ApJ...969..130L}.

We have considered only a very simple drag law with constant stopping time. A more general form of Epstein drag was given in \cite{2018MNRAS.480.2813H}, where the stopping time depends both on gas density and relative velocity between gas and dust. For the two instabilities discussed in this paper, the streaming instability and the settling instability, gas density perturbations are almost absent: a common approximation is that the gas is in fact incompressible \citep[e.g.][]{2005ApJ...620..459Y, 2018MNRAS.477.5011S}. Furthermore, for drift velocities that are highly subsonic, corrections to the stopping time due to the relative velocity are very small. For the streaming instability, it is definitely safe to take the stopping time to be constant, and for the settling instability as long as Stokes numbers remain much smaller than unity.

The thermodynamics of the disc was modeled in a simple way, assuming the gas is isothermal. More realistic models were considered in \cite{2023MNRAS.522.5892L}, who showed that in many cases, the DSI can be stabilized by vertical buoyancy, while new instabilities can arise as well. How these effects interaction with a dust size distribution remains to be explored.

Given that the contribution of the resonance to the backreaction integral is different for the three instabilities considered so far, it would be interesting to explore different resonant instabilities \citep{2018MNRAS.477.5011S} with a size distribution. The three instabilities studied here lead to three different categories: 
\begin{itemize}
\item{The resonance gives a negative contribution to the growth rate (SI).}
\item{The resonance gives a positive contribution to the growth rate (DSI). A wide size distribution makes the perturbation regular rather than singular.}
\item{The contribution of the resonance can be positive or negative, depending on the size distribution (\adi, see Paper I). The resulting perturbation is regular for wide size distributions.}
\end{itemize}
Apart from the double resonant angle for the DSI, even very narrow size distributions around the resonance lead to non-resonant behaviour, typically only for $\Delta\stokes/\stokes_0 \lesssim \sqrt{\mu}$ can we expect to get close to the monodisperse result. 

In conclusion, for realistically wide size distributions, dust-gas instabilities will mostly show non-resonant growth in the limit $\mu \ll 1$. The classical streaming instability does not survive at all in this limit, as the resonance has a stabilizing effect. The settling instability does survive, with growth rates that are very similar to the monodisperse case, but with a stronger dependence on $\mu$ compared to monodisperse results. \\

\noindent {\tiny \textit{Data Availability.} All data used to create figures in this work are available at \url{https://doi.org/10.4121/810f163a-b786-4366-ad9b-5794d71c4ede}}

\begin{acknowledgements}
We want to thank the anonymous referee for an insightful report. This project has received funding from the European Research Council (ERC) under the
European Union’s Horizon Europe research and innovation programme (Grant Agreement No. 101054502). This work made use of several open-source software packages. We acknowledge \texttt{numpy} \citep{Harris2020Natur.585..357H}, \texttt{matplotlib} \citep{Hunter2007CSE.....9...90H}, \texttt{scipy} \citep{Virtanen2020NatMe..17..261V}, and \texttt{psitools} \citep{2021MNRAS.502.1469M}. We acknowledge 4TU.ResearchData for supporting open access to research data.
\end{acknowledgements}

\bibliographystyle{aa} 
\bibliography{rdi} 

\begin{appendix}

\section{Governing equations for the inertial wave RDI}
\label{sec:appendix_eq}

We start from a standard stratified shearing box; equations (\ref{eq:dustcont})-(\ref{eq:gasmom}) with the appropriate shearing box accelerations
\begin{align}
    \bm{\alpha}_{\rm g} = 2\eta {\bf \hat x} - 2\bm{\Omega}\times {\bf{v}_{\rm g}} - \nabla\Phi_{\rm tot},\\
    \bm{\alpha}_{\rm d} = - 2\bm{\Omega}\times {\bf{v}_{\rm g}} - \nabla\Phi_{\rm tot},
\end{align}
where the potential $\Phi_{\rm tot} = -S\Omega x^2 + \Omega^2 z^2/2$ ($S$ is the shear rate of the disc), and the term involving $\eta$ is representing a global radial pressure gradient in the disc:
\begin{align}
\partial_t{\rho_{\rm g}} + \nabla\cdot ({\rho_{\rm g}} {\bf{v}_{\rm g}}) =& 0\, ,\\
\partial_t{\bf{v}_{\rm g}} + ({\bf{v}_{\rm g}}\cdot\nabla){\bf{v}_{\rm g}} =& 2\eta {\bf \hat x} -\frac{\nabla p}{{\rho_{\rm g}}} + \frac{1}{{\rho_{\rm g}}}\int \sigma \frac{{\bf u}-{\bf v}_{\rm g}}{\tau_{\rm s}}\rmd \tau_{\rm s} \nonumber\\
-& 2\bm{\Omega}\times {\bf v}_{\rm g} - \nabla\Phi_{\rm tot}\, ,\\
\partial_t\sigma + \nabla\cdot(\sigma {\bf{u}} )=& 0,\\
\partial_t{\bf{u}}
+ ({\bf{u}}\cdot\nabla){\bf{u}}
=&
-\nabla\Phi_{\rm tot}-2\bm{\Omega}\times {\bf u} - \frac{{\bf u}-{\bf v}_{\rm g}}{\tau_{\rm s}}.
\end{align}
 The equation of state for the gas is taken to be isothermal as usual: $p=c_{\rm g}^2\rho_{\rm g}$.

Consider a domain centered on $z=-z_0<0$\footnote{This choice makes the final result consistent with \cite{2020MNRAS.497.2715K}}, with vertical extent $L_z \ll z_0$, so that we can take the vertical gravitational acceleration to be constant:
\begin{align}
    \Phi_{\rm tot} = -S\Omega x^2 - \Omega^2 z_0 z = \Phi - \Omega^2 z_0 z.
\end{align}
From the vertical component of the dust momentum equation, we find that the equilibrium vertical drift is
\begin{align}
    \frac{u_z-{v}_{{\rm g},z}}{\tau_{\rm s}} = \Omega^2 z_0.
\end{align}
For the gas, in the vertical direction, hydrostatic equilibrium is assumed, with a horizontally uniform background density profile
\begin{align}
    \bar\rho_{\rm g}(z) = \rho_{\rm m} + \rho_{{\rm g},0}(z). 
\end{align}
Hydrostatic balance in the vertical direction then requires
\begin{align}
\frac{\nabla p}{{\rho_{\rm g}}} = - (1-\mu)\Omega^2z_0. 
\label{eq:hydro_equi}
\end{align}
Write $\rho_{\rm g}(t,x,y,z)=\bar \rho_{\rm g}(z) + \tilde \rho_{\rm g}(t,x,y,z)$:
\begin{align}
\partial_t\tilde\rho_{\rm g} + \nabla\cdot ((\bar\rho_{\rm g} + \tilde\rho_{\rm g}) {\bf{v}_{\rm g}}) =& 0\, ,\\
\partial_t{\bf{v}_{\rm g}} + ({\bf{v}_{\rm g}}\cdot\nabla){\bf{v}_{\rm g}} =& 2\eta {\bf \hat x} -\frac{c^2\nabla \tilde\rho_{\rm g}}{\bar \rho_{\rm g}+\tilde\rho_{\rm g}} +\frac{\tilde\rho_{\rm g}}{\bar \rho_{\rm g}+\tilde\rho_{\rm g}}\Omega^2z_0 {\bf\hat z}\nonumber\\
+& \frac{1}{\bar \rho_{\rm g}+\tilde\rho_{\rm g}}\int \sigma \frac{{\bf u}-{\bf v}_{\rm g}}{\tau_{\rm s}}\rmd \tau_{\rm s} \nonumber\\
-& 2\bm{\Omega}\times {\bf v}_{\rm g}-\nabla\Phi-\mu\Omega^2z_0{\bf\hat z}\, ,\\
\partial_t\sigma + \nabla\cdot(\sigma {\bf{u}} )=& 0,\\
\partial_t{\bf{u}}
+ ({\bf{u}}\cdot\nabla){\bf{u}}
=&
\Omega^2z_0-2\bm{\Omega}\times {\bf u} - \frac{{\bf u}-{\bf v}_{\rm g}}{\tau_{\rm s}}.
\end{align}
Note that the last term in the gas momentum equation is due to the term proportional to $\mu$ in (\ref{eq:hydro_equi}). Scale the gas momentum equation by choosing a typical velocity magnitude $V$, so that $|{\bf v}_{\rm g}|/V$ is of order unity:
\begin{align}
\partial_{\hat t}\hat {\bf v}_g 
+ (\hat {\bf v}_{\rm g}\cdot\hat\nabla)\hat{\bf v}_{\rm g} 
=& \frac{2\eta L_z}{V^2} {\bf \hat x} -\frac{c^2}{V^2}\frac{\hat\nabla \tilde\rho_{\rm g}}{\bar \rho_{\rm g}+\tilde\rho_{\rm g}} +\frac{\tilde\rho_{\rm g}}{\bar \rho_{\rm g}+\tilde\rho_{\rm g}}\frac{\Omega^2z_0 L_z}{V^2} \hat{\bf z}\nonumber\\
+& \frac{L_z}{V}\frac{1}{\bar \rho_{\rm g}+\tilde\rho_{\rm g}}\int \sigma \frac{\hat{\bf u}-\hat{\bf v}_{\rm g}}{\tau_{\rm s}}\rmd \tau_{\rm s} \nonumber\\
-& \frac{L_z}{V}2\bm{\Omega}\times \hat{\bf v}_{\rm g} -\frac{\hat\nabla\Phi}{V^2} - \frac{\mu \Omega^2 z_0 L_z}{V^2}{\bf \hat z}\, ,
\end{align}
where $\hat {\bf v}_{\rm g}={\bf{v}_{\rm g}}/V$, and length and time scales are normalized by $L_z$ and $L_z/V$, respectively. 
The left hand side is explicitly of order unity. Consider the terms on the right hand side operating in the vertical direction: pressure, buoyancy, drag, and the pressure correction (second, third, fourth, and last term on the right hand side, respectively). Define a small parameter
\begin{align}
    \epsilon \equiv \frac{\Omega^2z_0^2}{c^2}\frac{L_z}{z_0}.
\end{align}
This is a measure for the relative gas density variation over the vertical domain, and if $L_z \ll z_0$, as it must be, we have that $\epsilon \ll 1$ as long as $z_0 \sim H$. In standard Boussinesq analysis, one assumes that the density variations are due to mixing and are therefore 
\begin{align}
\frac{\tilde\rho_{\rm g}}{\rho_m} = O\left(\epsilon\right) \ll 1.
\label{eq:bous_mix}
\end{align}
This means that the buoyancy term can be neglected compared to the pressure term. Note that this is a consequence of the isothermal equation of state. By choosing a velocity scale $V=\Omega^2z_0\tau_{\rm ave}$, it is possible to make the drag term and the pressure term of the same magnitude (for $\mu \sim 1$). Note that this implies that we expect all gas motions to be very subsonic for $\Omega z_0 \sim c_{\rm g}$.

The non-dimensional version of the continuity equation is
\begin{align}
\frac{\partial_{\hat t}\tilde\rho_{\rm g}}{\rho_{\rm m}} 
+ \hat{\bf v}_{\rm g}\cdot \tilde\nabla\left(\frac{\bar\rho_{\rm g} + \tilde\rho_{\rm g}}{\rho_{\rm m}}\right) 
+ \frac{\bar\rho_{\rm g} + \tilde\rho_{\rm g}}{\rho_{\rm m}}\hat\nabla\cdot \hat{\bf v}_{\rm g}
= 0,
\end{align}
and because of (\ref{eq:bous_mix}) and the fact that $\hat\nabla\bar\rho_{\rm g} \sim \epsilon \rho_{\rm m}$, we can see that to lowest order in $\epsilon$, the gas follows the incompressibility condition, which is, in its dimensional form
\begin{align}
\nabla\cdot{\bf{v}_{\rm g}}=0.
\end{align}
The dimensional gas momentum equation is
\begin{align}
\partial_t{\bf{v}_{\rm g}} + ({\bf{v}_{\rm g}}\cdot\nabla){\bf{v}_{\rm g}} =& 2\eta {\bf \hat x} -\frac{\nabla p}{\rho_{\rm g}} - 2\bm{\Omega}\times {\bf v}_{\rm g}-\nabla\Phi\nonumber\\
+& \frac{1}{\rho_{\rm g}}\int \sigma \frac{{\bf u}-{\bf v}_{\rm g}}{\tau_{\rm s}}\rmd \tau_{\rm s} - \mu\Omega^2z_0, 
\end{align}
with the understanding that the background pressure gradient does not feature in $p$. These are the equations that are used for analysis of the SI (with $z_0=0$), and the settling instability for $z_0 \neq 0$, in which case our equations are identical to those of \cite{2020MNRAS.497.2715K}. In addition, we have established that we expect the gas to behave as an incompressible fluid, also for $z_0 \neq 0$.

\section{Dispersion relation for the inertial wave RDI}
\label{sec:appendix_disp}

Here, we follow the procedure of \cite{2021MNRAS.502.1579P} to get the dispersion relation in matrix form, but with an added vertical velocity component. We can obtain an expression for the dispersion relation by eliminating all quantities except ${\bf{\hat{v}}_{\rm g}}$. Starting from equations (\ref{eq:eig_gasdens})-(\ref{eq:eig_dustvel}), first eliminate the gas density perturbation through (\ref{eq:eig_gasdens}):
\begin{align}
\frac{{\hat{\rho}_{\rm g}}}{{\rho_{\rm g}^{(0)}}} = \frac{{\bf k}\cdot {\bf{\hat{v}}_{\rm g}}}{\omega - k_x {v^{(0)}_{{\rm g}x}}},
\end{align}
and write the gas momentum equation as
\begin{eqnarray}
\mathsf{P} {\bf{\hat{v}}_{\rm g}}
+ \frac{\rmi}{{\rho_{\rm g}^{(0)}}}\int \hat \sigma\frac{\Delta{\bf u}^{(0)}}{\taus}\rmd\taus
+ \frac{\rmi}{{\rho_{\rm g}^{(0)}}}\int \sigma^{(0)}\frac{{\bf \hat u}-{\bf{\hat{v}}_{\rm g}}}{\taus}\rmd\taus
=0\label{eq:gasmomP},
\end{eqnarray}
where the matrix $\mathsf{P}$ is given by
\begin{eqnarray}
\mathsf{P} = \left(\begin{array}{ccc}
- \omega_{\rm g}  + \frac{k_x^2c^2}{\omega_{\rm g}} & 2\rmi\Omega &  \frac{k_xk_zc^2}{\omega_{\rm g}} \\
-\rmi (2\Omega - S) & - \omega_{\rm g} & 0\\
\frac{k_xk_zc^2}{\omega_{\rm g}}  & 0 & - \omega_{\rm g} + \frac{k_z^2c^2}{\omega_{\rm g}}
\end{array}\right),
\end{eqnarray}
with shifted frequency $\omega_{\rm g} = \omega -k_x{v^{(0)}_{{\rm g}x}}$. The gas drag terms in the gas momentum equation (\ref{eq:gasmomP}) read:
\begin{align}
\frac{\rmi}{{\rho_{\rm g}^{(0)}}}\int \frac{\sigma^{(0)}}{\taus} & \left[\Delta{\bf u}^{(0)} \frac{\hat\sigma}{\sigma^{(0)}} + {\bf \hat u} - {\bf{\hat{v}}_{\rm g}}\right]\rmd\taus
=\nonumber\\
& \int \mathcal{K}(\taus)\left[\Delta{\bf u}^{(0)} \frac{{\bf k}\cdot {\bf\hat u}}{\omega-k_xu_x^{(0)} -k_zu_z^{(0)}} + {\bf \hat u} - {\bf{\hat{v}}_{\rm g}}\right]\rmd\taus,
\label{eq:gasdragint}
\end{align}
with kernel $\mathcal{K}= \rmi \sigma^{(0)}/({\rho_{\rm g}^{(0)}}\taus)$ and we have used the dust continuity equation to write $\hat\sigma$ in terms of ${\bf \hat u}$. If we define a matrix $\mathsf{V}$ such that
\begin{align}
\int \mathcal{K}(\taus)\left[\Delta{\bf u}^{(0)} \frac{\hat\sigma}{\sigma^{(0)}} + {\bf \hat u} - {\bf{\hat{v}}_{\rm g}}\right]\rmd\taus
=
\int \mathcal{K}(\taus)\left[\mathsf{V}(\taus){\bf \hat u} - {\bf{\hat{v}}_{\rm g}}\right]\rmd\taus.\label{eq:gasmomV}
\end{align}
It is easily verified that we need
\begin{align}
\mathsf{V} = \mathsf{I} + \frac{1}{\omega-k_xu_x^{(0)}  -k_z u_z^{(0)}}\left(\begin{array}{ccc}
\Delta u_x^{(0)} k_x & 0 & \Delta u_x^{(0)} k_z\\
\Delta u_y^{(0)} k_x & 0 & \Delta u_y^{(0)} k_z\\
\Delta u_z^{(0)} k_x & 0 & \Delta u_z^{(0)} k_z
\end{array}\right).
\end{align}
We want to get an expression for ${\bf \hat u}$ in terms of ${\bf{\hat{v}}_{\rm g}}$. The dust momentum equation (\ref{eq:eig_dustvel}) gives, after eliminating gas density:
\begin{align}
\left(k_x u_x^{(0)} + k_zu_z^{(0)} -\omega-\frac{\rmi}{\taus}\right) {\bf \hat u}
+ \rmi S\hat u_{x}{\bf \hat y}
&-2\rmi\bm{\Omega}\times {\bf \hat u}
=\nonumber\\
&\rmi \frac{{\bf k}\cdot {\bf{\hat{v}}_{\rm g}}}{\omega_{\rm g}}\frac{\Delta {\bf u}^{(0)}}{\taus}
-\rmi \frac{{\bf{\hat{v}}_{\rm g}}}{\tau_s}.
\end{align}
Write as matrix equation
\begin{align}
\mathsf{A}(\taus){\bf\hat u} = \mathsf{D}(\taus) {\bf{\hat{v}}_{\rm g}},
\end{align}
with
\begin{align}
\mathsf{A}=\left(\begin{array}{ccc}
d & 2\rmi\Omega & 0\\
\rmi(S-2\Omega) & d & 0\\
0 & 0 & d
\end{array}\right),
\end{align}
with $d = k_xu_x^{(0)} +k_zu_z^{(0)} - \omega - \rmi/\taus$, and
\begin{align}
\mathsf{D} = -\frac{\rmi}{\taus}\mathsf{I} + \frac{\rmi}{\taus\omega_{\rm g}}\left(\begin{array}{ccc}
\Delta u_x^{(0)} k_x & 0 & \Delta u_x^{(0)} k_z\\
\Delta u_y^{(0)} k_x & 0 & \Delta u_y^{(0)} k_z\\
\Delta u_z^{(0)} k_x & 0 & \Delta u_z^{(0)} k_z
\end{array}\right).
\end{align}
Hence ${\bf \hat u} = \mathsf{A}^{-1}\mathsf{D} {\bf{\hat{v}}_{\rm g}}$, which we can use in (\ref{eq:gasmomV}) to obtain
\begin{align}
\int \mathcal{K}(\taus) & \left[\Delta{\bf u}^{(0)} \frac{\hat\sigma}{\sigma^{(0)}} + {\bf \hat u} - {\bf{\hat{v}}_{\rm g}}\right]\rmd\taus
=\nonumber\\
& \int \mathcal{K}(\taus)\left[\mathsf{V}(\taus)\mathsf{A}^{-1}(\taus)\mathsf{D}(\taus) - \mathsf{I}\right]\rmd\taus  {\bf{\hat{v}}_{\rm g}} \equiv \mathsf{M} {\bf{\hat{v}}_{\rm g}}.
\label{eq:gasmomM}
\end{align}
The inverse of $\mathsf{A}$ is straightforward to calculate:
\begin{eqnarray}
\mathsf{A}^{-1}=\left(\begin{array}{ccc}
-\frac{d}{\kappa^2-d^2} & \frac{2\rmi\Omega}{\kappa^2-d^2} & 0\\
\frac{\rmi(S-2\Omega)}{\kappa^2-d^2} & -\frac{d}{\kappa^2-d^2} & 0\\
0 & 0 & \frac{1}{d}
\end{array}\right).
\end{eqnarray}
The dispersion relation is found by plugging (\ref{eq:gasmomM}) into (\ref{eq:gasmomP}) and is given by 
\begin{align}
\det(\mathsf{P} + \mathsf{M})=0.
\label{eq:dsi_dispersion}
\end{align}
This equation can be solved for the eigenvalues $\omega$ with the same tools as presented in \cite{2021MNRAS.502.1579P}.

\end{appendix}
\end{document}